\documentclass[smallextended,11pt,psfig,a4]{amsart}

\usepackage[latin1]{inputenc}
\usepackage{graphicx,amsmath,dsfont}
\usepackage{indentfirst}
\usepackage{amssymb}
\usepackage{amsfonts}
\usepackage{graphicx}
\usepackage{array}
\usepackage{amsthm}
\usepackage{float}
\usepackage{multirow}
\usepackage[normalem]{ulem}
\usepackage{braket}
\usepackage{lastpage}
\usepackage{subcaption}
\usepackage{indentfirst}
\usepackage{microtype}
\usepackage{fancyhdr}
\usepackage[all]{xy}
\usepackage{blkarray, bigstrut}
\usepackage[toc,page]{appendix}
\usepackage{amsmath}               
\usepackage{amsfonts}              
\usepackage{amsthm}                
\usepackage{esint}
\usepackage{graphics}
\usepackage{verbatim}
\usepackage{tikz}
\usepackage{hyperref}

\usepackage[textwidth = 155mm]{geometry}
\usetikzlibrary{automata, arrows}

\author[lv]{Newton Loebens}
\address{Newton Loebens, Instituto Federal de Mato Grosso do Sul. Aquidauana, MS 79200-000 Brazil. Telephone:(67) 3240-1600.}
\email{newtonloebens@gmail.com}

\def\bma{\begin{bmatrix}}
\def\ema{\end{bmatrix}}
\def\bex{\begin{example}}
\def\eex{\end{example}}
\def\beq{\begin{equation}}
\def\eeq{\end{equation}}

\def\ov{\overline}

\theoremstyle{plain}
\newtheorem{teo}{Theorem}[section]

\newtheorem{corollary}[teo]{Corollary}
\newtheorem{defi}[teo]{Definition}
\newtheorem{prop}[teo]{Proposition}

\theoremstyle{definition}
\newtheorem{remark}{Remark}

\begin{document}
\pagestyle{plain}


\title{Continuous-time open quantum walks in one dimension: matrix-valued orthogonal polynomials \\ and Lindblad generators}

\maketitle

\begin{abstract}
We study continuous-time open quantum walks in one dimension through a matrix representation, focusing on nearest-neighbor transitions for which an associated weight matrix exists. Statistics such as site recurrence are studied in terms of matrix-valued orthogonal polynomials and explicit calculations are obtained for classes of Lindblad generators that model quantum versions of birth-death processes. Emphasis is given to the technical distinction between the cases of a finite or infinite number of vertices. Recent results for open quantum walks are adapted in order to apply the folding trick to continuous-time birth-death chains on the integers. Finally, we investigate the matrix-valued Stieltjes transform associated to the weights.
\end{abstract}

\medskip

\textbf{Keywords}:\\
Continuous-time open quantum walks.
Matrix-valued orthogonal polynomials.
Stieltjes transform.
Lindblad generator.
Matrix representation.

\tableofcontents

\section{Introduction}{\color{black}
Random walks have been a fundamental concept in the study of stochastic processes and probability theory for many decades \cite{aliyev,KMc2,sandric}. In the field of quantum mechanics, the concept of quantum walks has emerged as a robust tool for exploring quantum systems' behavior and dynamics \cite{attal,portugal,vene}. Quantum walks can be categorized into various types, and one particularly intriguing category is Open Quantum Walks (OQWs) \cite{attal}. This process introduces the influence of the environment, which leads to a richer set of dynamics and behaviors when compared to the classical random walks induced by Markov chains, making them an exciting area of research in quantum information and quantum computation. Continuous-Time Open Quantum Walks (CTOQWs) \cite{pele} represent a specific class of OQWs where the evolution of a quantum walker in a graph is continuous and influenced by an initial quantum state.

\medskip

Inspired by the classical Birth-Death Processes (BDPs), this article develops a generalization from the perspective of CTOQWs, exploring promising valuable insights into the behavior of quantum walkers in systems that exhibit birth and death processes. Introducing a matrix representation for the generator of a CTOQW, we apply the theory of matrix orthogonal polynomials to tridiagonal block matrices. The matrix orthogonal polynomial approach provides a powerful framework for analyzing the representation of those generators, enabling us to gain a deeper understanding of CTOQWs and their connection to BDPs. This technique has been applied in the case of unitary quantum walks, where the relevant orthogonal polynomials are described in terms of the theory of CMV matrices \cite{grcpam,cmv}. Regarding the setting of open quantum dynamics \cite{benatti,petruc,davies}, the problem of obtaining orthogonal polynomials and associated weights is an interesting one as well, although we would have to consider operators which are no longer unitary. A first step in this direction has been discussed in \cite{jl}, where a procedure for obtaining weight matrices associated with open quantum walks (OQWs) \cite{attal} on the half-line was described. In \cite{dILL} it was studied the case of discrete-time quantum Markov chains on the line, as defined by S. Gudder \cite{gudder}, and gave a collection of some nontrivial examples where the spectral representation can be explicitly achieved.

\medskip

Analogously to the references above, we can employ the matrix orthogonal polynomial framework to explore various statistical aspects of CTOQWs, including recurrence patterns and transition probabilities. We utilize the Stieltjes transform as a key tool to analyze these statistics, offering an effective method to understand the intricate dynamics of quantum walkers in quantum systems influenced by tridiagonal block matrices. We are particularly interested in the recurrence of CTOQWs in this work. A first step in this direction can be seen in \cite{loebens}.

\medskip

Let us recall the classical BDPs.} Birth-death processes on $\mathbb{Z}_{\geq0}$ are continuous-time Markov chains characterized by a set of birth-death rates $\{(\lambda_n,\mu_n), n\geq0\}$ such that $\lambda_n>0,n\geq0$, $\mu_n>0,n\geq1$ and $\mu_0\geq0$ (see \cite{ander}). The transition function $P(t)=(P_{ij}(t))$ satisfies the following conditions as $t\to0^+$:
$$
P_{ij}(t)=\begin{cases}
\lambda_it+o(t),&\mbox{if}\quad j=i+1,\\
\mu_it+o(t),&\mbox{if}\quad j=i-1,\\
1-(\lambda_i+\mu_i)t+o(t),&\mbox{if}\quad j=i.
\end{cases}
$$
The matrix corresponding to the infinitesimal operator associated with the process is given by
\begin{equation}\label{Pbirthinf}
\mathcal{A}=\begin{bmatrix}
       -(\lambda_0+\mu_0) &\lambda_0 & 0 &0& \cdots \\
      \mu_1 &  -(\lambda_1+\mu_1)  & \lambda_1 & 0&\cdots \\
      0 & \mu_2 &  -(\lambda_2+\mu_2)  & \lambda_2&\cdots \\
      \vdots & \vdots & \vdots & \ddots&\ddots
    \end{bmatrix}.
\end{equation}
Following the classical work of S. Karlin and J. McGregor \cite{KMc3, KMc2}, we can apply Favard's Theorem to the Jacobi matrix \eqref{Pbirthinf} and assure the existence of a probability spectral measure $\psi$ supported on $[0,\infty)$ associated with $\mathcal{A}$. Moreover, if we define the sequence of polynomials $\{Q_n(x)\}_{n\geq 0}$ by the three-term recurrence relation
\begin{equation*}\label{3thermcod}
\begin{split}
Q_0(x)&=1,\quad Q_{-1}(x)=0, \\
-xQ_n(x)&=\lambda_nQ_{n+1}(x)-(\lambda_n+\mu_n)Q_n(x)+\mu_nQ_{n-1}(x),\;\;n\geq0,
\end{split}
\end{equation*}
that is, $-xQ(x)=\mathcal{A}Q(x),$ where $Q(x)=(Q_0(x),Q_1(x),\ldots)^T,$ then we have that the polynomials $\{Q_n(x)\}_{n\geq 0}$ are orthogonal with respect to $\psi$. This provides the so-called \emph{Karlin-McGregor formula} which gives an integral representation of the probability of reaching vertex $j$ at time $t$ given that the process started at vertex $i$, i.e. $P_{ij}(t)$. This formula is given by
\begin{equation*}\label{Kmcontd}
P_{ij}(t)=\frac{\displaystyle\int_{0}^{\infty}e^{-xt}Q_i(x)Q_j(x)d\psi(x)}{\displaystyle\int_{0}^{\infty}Q_j^2(x)d\psi(x)}.
\end{equation*}

\medskip

The main purpose of this paper is to analyze the spectral representation of some continuous-time open quantum walks (CTOQWs) by using the basic theory of matrix-valued orthogonal polynomials. The theory of orthogonal polynomials can be applied to the open quantum walks through an appropriate matrix representation that rises from the ``vec" application, whose role is to stack a density matrix in a unique bigger vector, and then a reversion of this application is made after an application of the matrix representation(see \cite{dILL,jl}).

\medskip

This dynamic is described by a quantum Markov semigroup with a specific Lindblad generator and performs an evolution of the initial density operator. Roughly speaking, the state at instant $t$ can be described by a pair $(X_t,\rho_t)$ with $X_t$ being the position of the particle at time $t$ and $\rho_t$ is the density operator describing the internal degrees of freedom of the corresponding vertex. We concentrate our results on CTOQWs whose vertices have all the same internal degrees of freedom, thereby the operators that describe the Lindblad generator will be acting on the same Hilbert space, and the matrices that describe the probability transitions will be squares.

\medskip
{\color{black}
The main result of this work is Equation \eqref{newstieltjes}, which expresses a formula for the Stieltjes transform of a CTOQW on the integer line in terms of Stieltjes transforms on the integer half-line. This transform associates a weight with a real function, enables us to evaluate the recurrence of CTOQWs, and offers a method for the construction of matrix weights that influence the orthogonality of the polynomials.} We remark that this result is valid for any semigroup having a matrix representation of the form \eqref{PhiinZ}, thus the folding trick is not retained to CTOQWs. For instance, we can also apply those formulas to quasi-birth-and-death processes.

\medskip

In Section 2 we review the ``vec" representation for completely positive maps and the matrix representation for maps of the form $\Psi(\rho)=G\rho+\rho G^*,$ where $G$ is the part of the Lindblad generator which is not completely positive. In Section 3 we discuss the model of CTOQWs and present its matrix representation. In Section 4 we recall the concept of matrix-valued orthogonal polynomials and show how the recurrence of CTOQWs can be associated to the Stieltjes transform. Section 5 develops the matrix representation for CTOQWs in the integer line, leading to the main result of this work, Equation \eqref{newstieltjes}, which associates the weight matrix of the walk in the half-line with the walk on the integer line. Section 6 illustrates the results with examples, giving explicit probabilities for different classes of Lindblad generators. {\color{black}In Section 7, an appendix is dedicated to recalling properties related to the existence of a matrix weight associated to the Lindblad generator.}

\section{General Settings}

Let $\mathcal{H}$ be a separable Hilbert space with inner product $\langle\,\cdot\,|\,\cdot\,\rangle$, whose closed subspaces will be referred to as subspaces for short. The superscript ${}^*$ will denote the adjoint operator. The Banach algebra $\mathcal{B}(\mathcal{H})$ of bounded linear operators on $\mathcal{H}$ is the topological dual of its ideal $\mathcal{I}(\mathcal{H})$ of trace-class operators with trace norm
$$
\|\rho\|_1=\operatorname{\mathrm{Tr}}(|\rho|),
\qquad
|\rho|=\sqrt{\rho^*\rho},
$$
through the duality \cite[Lec. 6]{attal_lec}
\beq \label{eq:dual}
\langle \rho,X \rangle = \operatorname{\mathrm{Tr}}(\rho X),
\qquad
\rho\in\mathcal{I}(\mathcal{H}),
\qquad
X\in\mathcal{B}(\mathcal{H}).
\eeq
If $\dim\mathcal{H}=k<\infty$, then $\mathcal{B}(\mathcal{H})=\mathcal{I}(\mathcal{H})$ is identified with the set of square matrices of order $k$, denoted $M_k(\mathbb{C})$. The duality \eqref{eq:dual} yields a useful characterization of the positivity of an operator $\rho\in\mathcal{I}(\mathcal{H})$,
\begin{equation*}\label{eq:pos-dual}
\rho\in\mathcal{I}(\mathcal{H}):
\quad
\rho\ge0 \; \Leftrightarrow \; \operatorname{\mathrm{Tr}}(\rho X)\ge0,
\quad
\forall X\in\mathcal{B}(\mathcal{H}),
\quad
X\ge0,
\end{equation*}
and similarly for the positivity of $X\in\mathcal{B}(\mathcal{H})$. {\color{black} In this work, we assume that we have a quantum particle walking either on the integer line, the integer half-line, or on a finite segment}, that is, we have that the set of vertices $V$ is labeled by $\mathbb{Z}$, $\mathbb{Z}_{\geq0}$ or a finite set $\{0,1,\dots,N\}$, respectively. We will also call vertices sites. The state of the system is described by a column vector
\beq\nonumber
\rho = \begin{bmatrix} \rho_0 \\ \rho_1 \\ \rho_2 \\ \vdots \end{bmatrix},
\qquad \rho_i\in\mathcal{I}(\mathcal{H}),
\qquad \rho_i\ge0,
\qquad \sum_{i\in V}\operatorname{\mathrm{Tr}}(\rho_i)=1.
\eeq

\medskip

The vector representation $vec(A)$ of $A\in M_k(\mathbb{C})$, given by stacking together its rows, will be a useful tool. For instance,
\begin{equation*}
A = \begin{bmatrix} a_{11} & a_{12} \\ a_{21} & a_{22} \end{bmatrix}
\quad\Rightarrow\quad
vec(A):=\begin{bmatrix} a_{11} \\ a_{12} \\ a_{21} \\ a_{22}\end{bmatrix}.
\end{equation*}
{\color{black} Let $\ov{B}=[\ov{b}_{ij}]$ for $B=[{b}_{ij}],$ that is, the entries of $\ov{B}$ are the complex conjugate entries of $B.$ } The $vec$ mapping satisfies $vec(AXB^T)=(A\otimes B)\,vec(X)$ for any square matrices $A, B, X$, with $\otimes$ denoting the Kronecker product. In particular, $vec(BXB^*)=vec(BX\ov{B}^T)=(B\otimes \ov{B})\,vec(X)$,
from which we can obtain the \textbf{matrix representation} $\widehat\Phi$ for a completely positive (CP) map $\sum_i B_i\cdot B_i^*$ when the underlying Hilbert space $\mathcal{H}$ is finite-dimensional:
\begin{equation*}\label{matrep}
\widehat\Phi = \sum_{i} \lceil B_{i} \rceil,
\qquad \lceil B \rceil := B \otimes \ov{B}.
\end{equation*}
Here the operators $B_i$ are identified with some matrix representation. We have that $\lceil B \rceil^* = \lceil B^*\rceil $, where $B^*$ denotes the Hermitian transpose {\color{black}(also known as conjugate transpose)} of a matrix $B$. The same idea can be applied to maps of the form $\Psi(\rho)=G\rho+\rho G^*.$ On this case the map $\Psi$ has matrix representation
\begin{equation*}
\hat{\Psi}=G\otimes I+I\otimes\overline{G}.
\end{equation*}
For more details, we refer the reader to the reference \cite{HornJo-topics}.

\section{Continuous-time open quantum walks}

An operator \textbf{semigroup}\index{semigroup} $\mathcal{T}$ on a Hilbert space $\mathcal{H}$ is a family of bounded linear operators $(T_t)$ acting
on $\mathcal{H}$, $t\geq 0,$ such that
\begin{equation}\label{def-semi}\nonumber
T_tT_s=T_{t+s},\quad s,t\in\mathbb{R}^+,\quad T_0=I_\mathcal{H}.
\end{equation}
If $t\mapsto T_t$ is continuous for the operator norm of $\mathcal{H}$, then $\mathcal{T}$ is said to be \textbf{uniformly
continuous}.\index{uniformly continuous semigroup} This class of semigroups is characterized by the following result:

\begin{teo}[\cite{bratteli}, page 161]
The following assertions are equivalent for a semigroup $\mathcal{T}$ on $\mathcal{H}:$
\begin{enumerate}
  \item $\mathcal{T}$ is uniformly continuous;
  \item There exists a bounded operator $L$ on $\mathcal{H}$ such that
  \begin{equation}\label{defi-ger}\nonumber
  T_t=e^{tL},\quad t\in \mathbb{R}^+.
  \end{equation}
Further, if the conditions are satisfied, then
\begin{equation}\label{defi-gerlim}\nonumber
L=\lim_{t\rightarrow0^+}\frac{T_t-I_\mathcal{B}}{t}.
\end{equation}
\end{enumerate}

The operator $L$ is called the \textbf{generator} of $\mathcal{T}.$ \index{generator of $\mathcal{T}$}
\end{teo}

\medskip

A trace-preserving semigroup $\mathcal{T}:=(\mathcal{T}_t)_{t\geq 0}$ of CP maps acting on $\mathcal{I}_1(\mathcal{H}),$ set of trace-class operators on $\mathcal{H},$ is called a \textbf{Quantum Markov Semigroup} (QMS)\index{Quantum Markov Semigroup (QMS)} on  $\mathcal{I}_1(\mathcal{H}).$ When $\lim_{t\rightarrow 0}||\mathcal{T}_t- Id||=0,$ $\mathcal{T}$ has a generator $\mathcal{L}=\lim_{t\rightarrow0^+}(\mathcal{T}_t-\textrm{Id})/t$ (see \cite{Lind}), which is a bounded operator on $\mathcal{I}_1(\mathcal{H}),$ also known as \textbf{Lindblad operator}.

\medskip

We consider a finite or countable set of vertices $V$ and then take the composite system
$$
\mathcal{H}=\bigoplus_{i\in V}\mathfrak{h}_i,
$$
where each $\mathfrak{h}_i$ denotes a separable Hilbert space. The label $i\in V$ is interpreted as being the position of the walker and, when the walker is located at the vertex $i\in V$, its internal state is encoded in the space $\mathfrak{h}_i,$ describing the internal degrees of freedom of the particle when it is sitting at site $i\in V.$ Since we will be considering only examples with $\mathfrak{h}_i=\mathfrak{h}_j$ for all $i,j\in V,$ we let $\mathfrak{h}_i=\mathfrak{h}$ for every $i\in V.$

\medskip

The set of diagonal density operators acting on $\mathcal{H}$ will be denoted by
$$
\mathcal{D}=\left\{\sum_{i\in V}\rho(i)\otimes\ket{i}\bra{i}:\;\rho(i)=\rho(i)^*,\quad\rho(i)\geq 0,\quad\sum_{i\in V}\textmd{Tr}(\rho(i))=1\right\}.
$$
\medskip

\begin{defi}[\cite{pele}]
A \textbf{Continuous-time Open Quantum Walk}\index{Continuous-time Open Quantum Random Walk CTOQW} (CTOQW) is an uniformly continuous QMS on $\mathcal{I}_1(\mathcal{H})$ with Lindblad
operator of the form
\begin{eqnarray}\label{Lind}
\nonumber  \mathcal{\mathcal{L}}:\mathcal{I}_1(\mathcal{H}) &\rightarrow &  \mathcal{I}_1(\mathcal{H})\\
  \rho &\mapsto & -i[H,\rho]+\sum_{i,j\in V}\left(S_i^j\rho S_i^{j^*}-\frac{1}{2}\{S_i^{j*}S_i^j,\rho\}\right),
\end{eqnarray}
where, consistently with the notation, we write $S_i^j=R_i^j\otimes\ket{j}\bra{i}$ for bounded operators $R_i^j\in \mathcal{B}(\mathfrak{h}_i,\mathfrak{h}_j).$ Moreover, $H$ and $S_i^j$ are bounded operators on $\mathcal{H}$ of the form $H=\sum_{i\in V}H_i\otimes\ket{i}\bra{i},$ $H_i$ is self-adjoint on $\mathfrak{h}_i,$ $S_i^j$ is a bounded operator on $\mathcal{H}$ with $\sum_{i,j\in V}S_i^{j*}S_i^j$ converging in the strong sense. Also, $[A,B]\equiv AB-BA$ is the commutator between $A$ and $B$ and $\{A,B\}\equiv AB+BA$ is the anti-commutator between $A$ and $B.$
\end{defi}

Then, we have $\rho=\sum_{i\in V}\rho(i)\otimes\ket{i}\bra{i}\in
\mathcal{D},$ $e^{t\mathcal{L}}(\rho)=\mathcal{T}_t(\rho)=\sum_{i\in V}\rho_t(i)\otimes\ket{i}\bra{i},\forall t\geq 0,$ with
$$
\frac{d}{dt}\rho_t(i)=-i[H_i,\rho_t(i)]+\sum_{j\in V}\left(R_j^i\rho_t(j) R_j^{i^*}-\frac{1}{2}\{R_i^{j*}R_i^j,\rho_t(i)\}\right).
$$

An alternative way to rewrite \eqref{Lind} is given by equation (18.7) in \cite{bardet}:
\begin{equation}\label{alternativeLindblad}
\mathcal{L}(\rho)=\sum_{i\in V}\left(G_i\rho(i)+\rho(i)G_i^*+\sum_{j\in V}R_j^i\rho(j) R_j^{i*}\right)\otimes\ket{i}\bra{i},
\end{equation}
where
\begin{equation*}\label{Gi}
G_i=-iH_i-\frac{1}{2}\sum_{j\in V}R_i^{j*}R_i^j.
\end{equation*}

\medskip

Further, we will present the matrix representation for CTOQWs, and this will be done by taking the representation given in Equation \eqref{alternativeLindblad}.

\medskip

The label $i\in V$ represents the position of the walker and, when the walker is located at $i\in V,$ its internal state is encoded in $\mathfrak{h}_i,$ that is, $\mathfrak{h}_i$  describes the internal degrees of
freedom of the walker when it is at site $i\in V.$

Starting the walk on site $\ket{i}$ with initial density operator $\rho\in\mathcal{S}(\mathfrak{h}_i)=\sum_{i\in V}\rho(i)\ket{i}\bra{i},$ the quantum measurement of the position gives rise to a probability distribution $p_0$ on $V,$ such that
$$
p_0(i)=\mathbb{P}(\mbox{the quantum particle is in site}\;\ket{i})=\textmd{Tr}(\rho(i))
$$
and for evolution on time $t\geq 0,$
$$
p_t(i)=\mathbb{P}(\mbox{the quantum particle, at time }t,\mbox{ is in site}\;\ket{i})=\textmd{Tr}(\rho_{t}(i)),
$$
where
$$
e^{t\mathcal{L}}(\rho)=\sum_{i\in V}\rho_t(i)\otimes\ket{i}\bra{i}.
$$

The vector and matrix representation of states and CP maps may be easily adapted to CTOQWs. In fact, since any element of $\mathcal{I}_1(\mathcal{H})$ is block diagonal, when $\dim\mathcal{H}<\infty$, it may be represented by combining the vector representations of the finite diagonal blocks,
$$
\rho=\sum_{i\in V} \rho_i\otimes|i\rangle\langle i|
\quad\Rightarrow\quad
\overrightarrow{\rho}:=\begin{bmatrix} vec(\rho_1) \\ vec(\rho_2) \\ \vdots \end{bmatrix}.
$$
The CTOQW \eqref{alternativeLindblad} admits the {\bf block matrix representation}
\begin{equation*}\label{eq:mrOQW}
\overrightarrow{e^{t\mathcal{L}}(\rho)} =
e^{t\widehat{\mathcal{L}}}\,\overrightarrow{\rho},
\qquad
\widehat{\mathcal{L}} =
\begin{bmatrix}
G_0^\alpha+\lceil B_{00} \rceil  & \lceil B_{01} \rceil& \lceil B_{02}\rceil & \cdots
\\[2pt]
\lceil B_{10} \rceil & G_1^\alpha+\lceil B_{11} \rceil  & \lceil B_{12}\rceil& \cdots
\\
\lceil B_{20} \rceil &\lceil B_{21} \rceil & G_2^\alpha+\lceil B_{22} \rceil  &  \cdots\\
\vdots & \vdots&\vdots&\ddots
\end{bmatrix},
\end{equation*}
where
$$
G_i^\alpha=\left(-iH_i-\frac{1}{2}\sum_{j\in V}R_i^{j*}R_i^j\right)\otimes I+
I\otimes\left(i\overline{H}_i-\frac{1}{2}\sum_{j\in V}\overline{R_i^{j*}R_i^j}\right).
$$

\medskip

We will often identify the Lindblad generator $\mathcal{L}$ with its block matrix representation and omit the hat, as the usage of such object will be clear from the context. Also, we will sometimes write $X$ instead of $\lceil X\rceil$ in contexts where no confusion arises.

\medskip

It is worth noting that although the above definitions concern CTOQWs on general graphs, in this paper we will deal exclusively with the one-dimensional situation which we may also call the quantum birth-death process, and represent the generator by
\begin{equation*}\label{tpm}
\hat{\mathcal{L}}=
\begin{bmatrix}
  B_0 & C_1 & && \\
  A_0& B_1 & C_2  && \\
 &A_1& B_2 & C_3  & \\
  &  & \ddots  &\ddots&\ddots
\end{bmatrix},
\end{equation*}
for certain operators $A_i, B_i, C_i$, and the remaining operators being equal to zero. The above representation is for a quantum particle walking on the integer half-line $\mathbb{Z}_{\geq0}$, but we will also study examples acting on a finite set $\{0,1,\ldots,N\}$ or the integer line $\mathbb{Z}$.

\section{Matrix-valued orthogonal polynomials}

In this section we introduce the Karlin-McGregor Formula for CTOQW with set of vertices of the forms $V=\{0,1,2,\ldots,N\}$ and $V=\mathbb{Z}_+=\{0,1,2\ldots\}.$ Then we will be able to give a recurrence criterion for vertex $\ket{0}$ based on the Stieltjes transform of the associated weights.

\medskip

Following \cite{detteC}, we pick $d\in \{1,2,3,\ldots\},$ $(A_n)_{n\geq0}, \;(B_n)_{n\geq0},$ and $(C_n)_{n\geq1}$, such that the block tridiagonal matrix
\begin{equation}\label{CTQtridiZ+}
\hat{\mathcal{L}}=\begin{bmatrix}
                B_0 & C_1 &  &  &  \\
                A_0 & B_1 & C_2 &  &  \\
                 & A_1 & B_2 &  C_3&  \\
                 &  & \ddots & \ddots & \ddots
              \end{bmatrix}
\end{equation}
represents a Lindblad generator of a CTOQW $\Lambda.$ Then define recursively the associated matrix-valued polynomials from the matrix $\hat{\mathcal{L}}$ on \eqref{CTQtridiZ+} by
\begin{equation}\label{CT3thermZ+}
\begin{split}
   Q_0(x)=&I_d,\quad Q_{-1}(x)=0_d \\
    -xQ_n(x) =&  Q_{n+1}(x)A_n+Q_n(x)B_n+Q_{n-1}(x)C_n,\;\;n=0,1,2,\ldots,
\end{split}
\end{equation}
that is, $Q(x)=(Q_0(x),Q_1(x),\ldots)$ are solutions of the equation $-xQ(x)=Q(x)\hat{\mathcal{L}}.$ Here we denote $I_d$ and $0_d$ the identity and the null matrix of dimension $d\times d$.

\medskip

We recall that $\Lambda_t'=\hat{\mathcal{L}}\Lambda_t,$ where $\Lambda_t=e^{t\hat{\mathcal{L}}}$ and define the two-variable function
$$f(x,t)=Q(x)\Lambda_t,\;\;x\in \mathbb{C},\;t\in[0,\infty).$$
One has
$$
\frac{\partial f(x,t)}{\partial t}=Q(x)\Lambda_t'=Q(x)\hat{\mathcal{L}}\Lambda_t=-xQ(x)\Lambda_t=-xf(x,t),\quad f(x,0)=Q(x),
$$
whose solution is $f(x,t)=e^{-xt}Q(x).$ Hence $e^{-xt}Q(x)=Q(x)\Lambda_t.$ Componentwise,
\begin{equation}\label{compwisec}
e^{-xt}Q_i(x)=\sum_{k=0}^{\infty}Q_k(x)\Lambda_{ki}(t),
\end{equation}
where $\Lambda_{ki}(t)$ is the $(k,i)$-th block of $\Lambda(t)$.

\medskip

If there exists a weight matrix $\Sigma$ such that the matrix-valued polynomials $\{Q_n(x)\}_{n\geq0}$ are orthogonal with respect to $\Sigma,$ in the following sense
$$
\int Q_j^*(x)d\Sigma(x)Q_i(x)=\delta_{ji}F_i,\;\;\det(F_i)\neq 0,
$$
then multiplying on the left side of \eqref{compwisec} by $Q_j^*(x)$ and integrating with respect to $\Sigma$ we obtain
$$\int_{\mathbb{R}}e^{-xt}Q_j^*(x)d\Sigma(x)Q_i(x)=\int_{\mathbb{R}}Q_j^*(x)d\Sigma(x)Q_j(x)\Lambda_{ji}(t),$$
therefore for any $i,j\in V,$
we have the \textbf{Karlin-McGregor Formula for CTOQWs}:\index{Karlin-McGregor Formula for! CTOQWs}
\begin{equation}\label{kmatrix}
\Lambda_{ji}(t)=\left(\int Q_j^*(x)d\Sigma(x)Q_j(x)\right)^{-1}\left(\int e^{-xt}Q_j^*(x)d\Sigma(x)Q_i(x)\right),
\end{equation}
$\Lambda(t)=(\Lambda_{ji}(t))_{j,i=0,1,\ldots}.$ For more details about how to construct this formula see \cite{detteC}.

\medskip

Sometimes we will write \eqref{kmatrix} as
$$
\Lambda_{ji}(t)=\Pi_j\left(\int e^{-xt}Q_j^*(x)d\Sigma(x)Q_i(x)\right),\quad
\Pi_j:=\left(\int Q_j^*(x)d\Sigma(x)Q_j(x)\right)^{-1}.
$$
Let $p_{ji;\rho}(t)$ represent the probability of reaching site $\ket{j}$ at instant $t$, given that we started at site $\ket{i}$ with initial density $\rho$ concentrated at $i.$ Then
\begin{equation*}\label{KMcGgeneral}
p_{ji;\rho}(t)=\textmd{Tr}\left[\mbox{vec}^{-1}\left(\Lambda_{ji}(t)\mbox{vec}(\rho)\right)\right]=\textmd{Tr}\left[\mbox{vec}^{-1}\left(\Pi_j\int e^{-xt}Q_j^*(x)d\Sigma(x)Q_i(x)\mbox{vec}(\rho)\right)\right].
\end{equation*}

For simplicity, we write the transition probabilities by
\begin{equation*}
p_{ji;\rho}(t)=\textmd{Tr}\left[\Pi_j\int e^{-xt}Q_j^*(x)d\Sigma(x)Q_i(x)\rho\right]
\end{equation*}
in contexts where no confusion arises.

\medskip

Consider a CTOQW with set of vertices $V.$ Given $\ket{i}\in V$ and $\rho\in\mathcal{S}(\mathfrak{h}_i),$ we say that $\ket{i}$ is $\rho$-recurrent if
$$
\int_{0}^{\infty}p_{ii;\rho}(t)dt=\infty.
$$
When $\ket{i}$ is recurrent for all densities, then we say that $\ket{i}$ is recurrent. This concept is associated with the weight matrices by the following theorem.

\begin{teo}\label{TeoRecurrence}
Consider a tridiagonal CTOQW on $\mathbb{Z}_{\geq0}=\{0,1,2,\ldots\}$ and let $\Sigma$ be its associated weight matrix. Vertex $\ket{j}$ is $\rho$-recurrent if and only if
\begin{equation*}\label{RecMeasure}
\lim_{\lambda\rightarrow 0}\textmd{Tr}\left[\Pi_j\int_{\mathbb{C}}\frac{Q_j^*(x)d\Sigma(x)Q_i(x)}{\lambda+x}\rho\right]=\infty.
\end{equation*}
\begin{proof}
For each pair $i,j\in V$ we have
\begin{equation}\nonumber
\begin{split}
   \int_{0}^{\infty}p_{ji;\rho}(t)dt =& \lim_{\lambda\rightarrow 0}\int_{0}^{\infty}e^{-\lambda t}p_{ji;\rho}(t)dt\\
   =&\lim_{\lambda\rightarrow 0}\int_{0}^{\infty}e^{-\lambda t}\textmd{Tr}\left[\Pi_j\int_{\mathbb{C}} e^{-xt}
     Q_j^*(x)d\Sigma(x)Q_i(x)\rho\right]dt \\
     =& \lim_{\lambda\rightarrow 0}\textmd{Tr}\left[\Pi_j\int_{\mathbb{C}}\left( \int_{0}^{\infty}e^{-(\lambda+x)t}dt\right)Q_j^*(x)d\Sigma(x)Q_i(x)\rho\right]\\
     =& \lim_{\lambda\rightarrow 0}\textmd{Tr}\left[\Pi_j\int_{\mathbb{C}}\frac{Q_j^*(x)d\Sigma(x)Q_i(x)}{\lambda+x}\rho\right].
\end{split}
\end{equation}
\end{proof}
\end{teo}

We recall the \textbf{Stieltjes transform} associated to $\Sigma:$
$$
B(z,\Sigma)=\int_\mathbb{C}\frac{d\Sigma(x)}{z-x},
$$
thus we obtain the straightforward consequence of Theorem \ref{TeoRecurrence}:
\begin{corollary}\label{CorRecurrence}
Consider a tridiagonal CTOQW on $\mathbb{Z}_{\geq0}=\{0,1,2,\ldots\}$ and let $\Sigma$ be its associated weight matrix. Vertex $\ket{0}$ is $\rho$-recurrent if and only if
\begin{equation*}\label{RecMeasure}
-\lim_{z\rightarrow 0}\textmd{Tr}\left[\Pi_0 B(z,\Sigma)\rho\right]=\infty.
\end{equation*}
\end{corollary}

It is crucial to note that not all polynomials induced by block tridiagonal matrices are orthogonalizable under any matrix weight. The non-trivial nature of establishing orthogonality in this context necessitates a discerning criterion for the existence of such weights. Within the framework of Section \ref{apendice}, the appendix recalls a criterion for the orthogonality of polynomials induced by block tridiagonal matrices, and a precise expression for a specific type of weight.

\section{Walks on $\mathbb{Z}$: the folding trick} Consider the generator of a tridiagonal CTOQW on $\mathbb{Z}$, given by
\begin{equation}\label{PhiinZ}
\hat{\mathcal{L}}=\left\lceil \begin{array}{ccc|ccccc}
\ddots & \ddots &  &  &  &  &  &  \\
  \ddots & G^\alpha_{-2}+\lceil B_{-2}\rceil  & \lceil C_{-1}\rceil  &  &  &  &  &  \\
   & \lceil A_{-2}\rceil  &G^\alpha_{-1}+\lceil B_{-1}\rceil  & \lceil C_{0}\rceil  &  &  &  &  \\
  \hline
   &    & \lceil A_{-1}\rceil  & G^\alpha_{0}+\lceil B_{0}\rceil  & \lceil C_{1}\rceil  & & &  \\
    &  &    & \lceil A_{0}\rceil  & G^\alpha_{1}+\lceil B_{1}\rceil  & \lceil C_{2}\rceil  & & \\
    & & &    & \lceil A_{1}\rceil  & G^\alpha_{2}+\lceil B_{2}\rceil  & \lceil C_{3}\rceil  &  \\
     &  &  &  &  & \ddots & \ddots &\ddots
\end{array}\right\rceil ,
\end{equation}
where all blocks are matrices of order $d^2,$ thus $\dim(\mathfrak{h})=d.$

We assume that there exists a sequence of $d^2\times d^2$ Hermitian matrices $(E_n)_{n\in\mathbb{Z}}$ and non-singular matrices $(R_n)_{n\in\mathbb{Z}}$ such that
\begin{equation}\label{condiequiv}
\begin{split}
  \lceil A_n\rceil ^*R_{n+1}^*R_{n+1} &= R_{n}^*R_{n}\lceil C_{n+1}\rceil ,  \; n\geq0\\
  R_{-n-1}^*R_{-n-1}\lceil C_{-n}\rceil  &=  \lceil A_{-n-1}\rceil ^*R_{-n}^*R_{-n}, \; n\geq 0,
\end{split}
\qquad R_n(G^\alpha_{n}+\lceil B_{n}\rceil )=E_nR_n,\; n\in\mathbb{Z}.
\end{equation}
Let us define
$$
\Pi_j:=R_j^*R_j,\;\;j\in\mathbb{Z}.
$$
Consider the two independent families of matrix-valued polynomials defined recursively from \eqref{PhiinZ} as
\begin{equation}\label{polsZgeneralOQW}
\begin{split}
Q_0^1(x) &= I_{d^2}, \quad Q_0^2(x)=0_{d^2},\\
Q_{-1}^1(x) &= 0_{d^2},\quad Q_{-1}^2(x)=I_{d^2}, \\
-xQ_n^\alpha (x) &= Q_{n+1}^\alpha(x)\lceil A_n\rceil +Q_{n}^\alpha(x)(G^\alpha_{n}+\lceil B_{n}\rceil )+Q_{n-1}^\alpha(x)\lceil C_n\rceil ,\quad \alpha=1,2,\;\;n\in\mathbb{Z},
\end{split}
\end{equation}
where we have the block vector $Q^\alpha(x)=\left(\ldots,Q_{-2}^\alpha(x),Q_{-1}^\alpha(x),Q_{0}^\alpha(x),Q_{1}^\alpha(x),Q_{2}^\alpha(x),\ldots\right),$ $\alpha=1,2,$ satisfying $-xQ^\alpha(x)=Q^\alpha(x)\hat{\mathcal{L}}$.

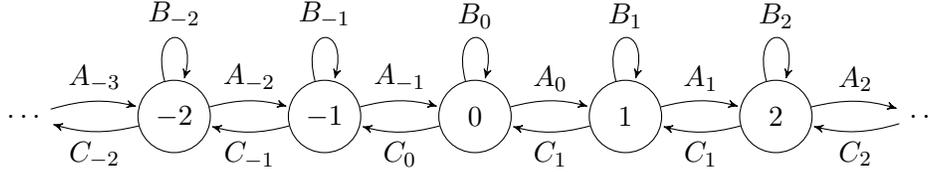
\begin{figure}[h!]
$$\begin{tikzpicture}[>=stealth',shorten >=1pt,auto,node distance=2cm]
\node[state] (q-2)      {$-2$};
\node[state]         (q-1) [right of=q-2]  {$-1$};
\node[state]         (q0) [right of=q-1]  {$0$};
\node[state]         (q1) [right of=q0]  {$1$};
\node[state]         (q2) [right of=q1]  {$2$};
\node         (qr) [right of=q2]  {$\ldots$};
\node         (ql) [left of=q-2]  {$\ldots$};
\path[->]          (ql)  edge         [bend left=15]  node[auto] {$A_{-3}$}     (q-2);
\path[->]          (q-2)  edge         [bend left=15]   node[auto] {$A_{-2}$}     (q-1);
\path[->]          (q-1)  edge         [bend left=15]   node[auto] {$A_{-1}$}     (q0);
\path[->]          (q0)  edge         [bend left=15]   node[auto] {$A_0$}     (q1);
\path[->]          (q1)  edge         [bend left=15]   node[auto] {$A_1$}     (q2);
\path[->]          (q2)  edge         [bend left=15]   node[auto] {$A_2$}     (qr);
\path[->]          (qr)  edge         [bend left=15] node[auto] {$C_{2}$}       (q2);
\path[->]          (q2)  edge         [bend left=15]   node[auto] {$C_{1}$}     (q1);
\path[->]          (q1)  edge         [bend left=15]  node[auto] {$C_{1}$}      (q0);
\path[->]          (q0)  edge         [bend left=15]   node[auto] {$C_{0}$}     (q-1);
\path[->]          (q-1)  edge         [bend left=15]  node[auto] {$C_{-1}$}      (q-2);
\path[->]          (q-2)  edge         [bend left=15]  node[auto] {$C_{-2}$}      (ql);
  \draw [->] (q1) to[loop above]node[auto] {$B_{1}$}  (q1);
   \draw [->] (q2) to[loop above]node[auto] {$B_{2}$}  (q2);
    \draw [->] (q-2) to[loop above] node[auto] {$B_{-2}$} (q-2);
     \draw [->] (q-1) to[loop above]node[auto] {$B_{-1}$}  (q-1);
      \draw [->] (q0) to[loop above]node[auto] {$B_{0}$}  (q0);
\end{tikzpicture}$$
\caption{Generator $\hat{\mathcal{L}}$ of a CTOQW on $\mathbb{Z}$.}
\label{originalQMConZ}
\end{figure}

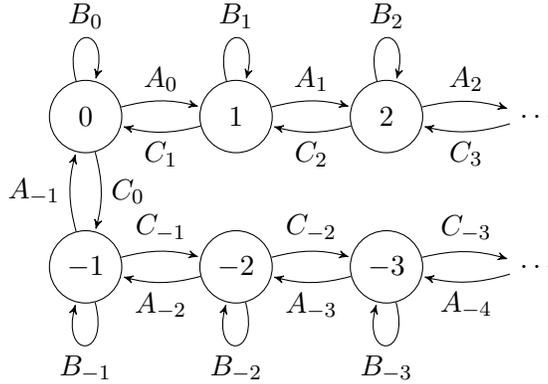
\begin{figure}[h]
$$\begin{tikzpicture}[>=stealth',shorten >=1pt,auto,node distance=2cm]
  \node[state] (q1)      {$0$};
  \node[state] (q2) [right of=q1]  {$1$};
  \node[state] (q3) [right of=q2]  {$2$};
  \node         (q4) [right of=q3]  {$\ldots$};
  \node[state] (q-1) [below of=q1]     {$-1$};
  \node[state] (q-2) [right of=q-1]  {$-2$};
  \node[state] (q-3) [right of=q-2]  {$-3$};
  \node         (q-4) [right of=q-3]  {$\ldots$};
  \path[->]          (q1)  edge         [bend left=15]   node[auto] {$A_0$}     (q2);
  \path[->]          (q2)  edge         [bend left=15]   node[auto] {$A_1$}     (q3);
  \path[->]          (q3)  edge         [bend left=15]   node[auto] {$A_2$}     (q4);
  \path[->]          (q2)  edge         [bend left=15]   node[auto] {$C_1$}     (q1);
  \path[->]          (q3)  edge         [bend left=15]   node[auto] {$C_2$}     (q2);
  \path[->]          (q4)  edge         [bend left=15]   node[auto] {$C_3$}     (q3);
  \draw [->]         (q1) to[loop above]  node[auto] {$B_0$} (q1);
  \draw [->]         (q2) to[loop above]  node[auto] {$B_1$}(q2);
  \draw [->]         (q3) to[loop above]  node[auto] {$B_2$}(q3);
  \path[->]          (q-1)  edge         [bend left=15]   node[auto] {$C_{-1}$}     (q-2);
  \path[->]          (q-2)  edge         [bend left=15]   node[auto] {$C_{-2}$}     (q-3);
  \path[->]          (q-3)  edge         [bend left=15]   node[auto] {$C_{-3}$}     (q-4);
  \path[->]          (q-2)  edge         [bend left=15]   node[auto] {$A_{-2}$}     (q-1);
  \path[->]          (q-3)  edge         [bend left=15]   node[auto] {$A_{-3}$}     (q-2);
  \path[->]          (q-4)  edge         [bend left=15]   node[auto] {$A_{-4}$}     (q-3);
    \draw [->]       (q-1) to[loop below]  node[auto] {$B_{-1}$} (q-1);
  \draw [->]         (q-2) to[loop below]  node[auto] {$B_{-2}$}(q-2);
  \draw [->]         (q-3) to[loop below]  node[auto] {$B_{-3}$}(q-3);
  \path[->]          (q1)  edge         [bend left=15]   node[auto] {$C_0$}     (q-1);
  \path[->]          (q-1)  edge         [bend left=15]   node[auto] {$A_{-1}$}     (q1);
\end{tikzpicture}$$
\caption{Folded walk of $\hat{\mathcal{L}}$ on $\mathbb{Z}_{\geq0}\times\{1,2\}$ given by $\breve{\mathcal{L}}.$}
\label{foldingQMC}
\end{figure}

As in the classical case, we introduce the block tridiagonal matrix
\begin{equation*}\label{Phifold}
\breve{\mathcal{L}}=
\begin{bmatrix}
  D_0 & N_1 &  & & \\
  M_0&D_1 & N_2   &&  \\
  &M_1&D_2& N_3    & \\
   &  & \ddots &\ddots &\ddots
\end{bmatrix},
\end{equation*}
where each block entry is a $2d^2\times 2d^2$ matrix, given by
$$
\begin{array}{rlrll}
  D_0=&\begin{bmatrix}
        G^\alpha_{0}+\lceil B_{0}\rceil  & \lceil A_{-1}\rceil \\
       \lceil C_0\rceil  & G^\alpha_{-1}+\lceil B_{-1}\rceil
      \end{bmatrix}, &
  M_n=&\begin{bmatrix}
        \lceil A_{n}\rceil  & 0 \\
        0 & \lceil C_{-n-1}\rceil
      \end{bmatrix},&n\geq 0, \\
   D_n=&\begin{bmatrix}
        G^\alpha_{n}+\lceil B_{n}\rceil  & 0 \\
        0 & G^\alpha_{-n-1}+\lceil B_{-n-1}\rceil
      \end{bmatrix}, &
   N_n=&\begin{bmatrix}
        \lceil C_{n}\rceil  & 0 \\
        0 & \lceil A_{-n-1}\rceil
      \end{bmatrix},&n\geq 1.
\end{array}
$$
The term \emph{folding trick} comes from the transformation of the original generator $\hat{\mathcal{L}}$, whose graph is represented in Figure \ref{originalQMConZ},
to the generator described by $\breve{\mathcal{L}},$ which is represented by the folded walk in Figure \ref{foldingQMC}.

Note that $\breve{\mathcal{L}}$ is a block tridiagonal matrix on $\mathbb{Z}_{\geq0},$ thereby we can apply all the properties we have seen in previous sections. The following $2d^2\times 2d^2$ matrix polynomials are defined in terms of \eqref{polsZgeneralOQW},
\begin{equation}\label{blobloQn}
\mathcal{Q}_n(x)=
\begin{bmatrix}
  Q_{n}^1(x) &Q_{-n-1}^1(x) \\
    Q_{n}^2(x)& Q_{-n-1}^2(x)
\end{bmatrix},\;\;n\geq 0,
\end{equation}
and these satisfy
\begin{align*}
x\mathcal{Q}_{0}(x) =& \mathcal{Q}_{1}(x)M_0+\mathcal{Q}_{0}(x)D_0,\;\; \mathcal{Q}_0(x)=I_{2d^2}, \\
x\mathcal{Q}_{n}(x) =& \mathcal{Q}_{n+1}(x)M_n+\mathcal{Q}_{n}(x)D_n+\mathcal{Q}_{n-1}(x)N_n,\;\;n=1,2,\ldots
\end{align*}
The leading coefficient of $\mathcal{Q}_{n}(x)$ is always a nonsingular matrix. Moreover, for
$$
\breve{R}_n:=\begin{bmatrix}
       R_n & 0_{d^2} \\
       0_{d^2} & R_{-n-1}
     \end{bmatrix},\;n\geq 0,\quad
\breve{E}_0:=\begin{bmatrix}
               E_0 & R_0\lceil A_{-1}\rceil R_{-1}^{-1} \\
               R_{-1}\lceil C_0\rceil R_0^{-1} & E_{-1}
             \end{bmatrix},\quad
\breve{E}_n:=\begin{bmatrix}
               E_n & 0_{d^2} \\
               0_{d^2} & E_{-n-1}
             \end{bmatrix},\;n\geq 1,
$$
we see that the block matrices of $\breve{\mathcal{L}}$ satisfy the conditions \eqref{condiequiv} for $n\geq 0:$
\begin{equation*}\label{condiequivbreve}
M_{n}^*\breve{R}_{n+1}^*\breve{R}_{n+1}  =  \breve{R}_{n}^*\breve{R}_{n}N_{n+1}, \;\;\;   \breve{R}_nD_n=\breve{E}_n\breve{R}_n,
\end{equation*}
where matrices $\breve{R}_{n}$ are non-singular and $\breve{E}_n$ are Hermitian for all $n\geq 0$. Defining
$$
\breve{\Pi}_j:=\breve{R}_j^*\breve{R}_j\in M_{2d^2}(\mathbb{C}),\;\;j=0,1,2,\ldots,
$$
the correspondence between $\breve{\Pi}_j$ and $\Pi_j$ is
$$
\breve{\Pi}_j:=\begin{bmatrix}
                 \Pi_j & 0_{d^2} \\
                 0_{d^2} & \Pi_{-j-1}
               \end{bmatrix}.
$$
By \cite{detteC}, there exists a weight matrix $W$ leading to the Karlin-McGregor formula for $\breve{\Lambda}=e^{t\breve{\mathcal{L}}}:$
\begin{equation}\label{KMFqmc}
\breve{\Lambda}_{ji}(t)=\breve{\Pi}_j\int_\mathbb{R} e^{-xt}\mathcal{Q}_{j}^*(x)dW(x)\mathcal{Q}_{i}(x).
\end{equation}
Once we have found the weight matrix appearing on \eqref{KMFqmc}, we can also obtain the blocks $\Lambda_{ji}(t)$ of the original walk generated by $\hat{\mathcal{L}}.$ The key for this operation is the following proposition:

\begin{prop}\label{Phibvsn} Assume that $\hat{\mathcal{L}}$ is the generator of a CTOQW of the form \eqref{PhiinZ}.
The relation between $\breve{\Lambda}_{ji}(t)$ and $\Lambda_{ji}(t)$ is
\begin{equation}\label{PHIbrevevshat}
\breve{\Lambda}_{ji}(t)=\begin{bmatrix}
               \Lambda_{ji}(t) & \Lambda_{j,-i-1}(t) \\
               \Lambda_{-j-1,i}(t) & \Lambda_{-j-1,-i-1}(t)
             \end{bmatrix},\;i,j\in\mathbb{Z}_{\geq0}.
\end{equation}
\begin{proof}
First we use \cite[Proposition 7.1]{dILL} (replace $\breve{\Phi}_{ji}^{(n)}$ and $\hat{\Phi}_{ji}^{(n)}$ by $\breve{\mathcal{L}}^n_{ji}$ and $\hat{\mathcal{L}}^n_{ji}$ respectively) to obtain that
$$
\breve{\mathcal{L}}_{ji}^{n}=\begin{bmatrix}
              \hat{\mathcal{L}}_{ji}^{n} & \hat{\mathcal{L}}_{j,-i-1}^{n} \\
               \hat{\mathcal{L}}_{-j-1,i}^{n} & \hat{\mathcal{L}}_{-j-1,-i-1}^{n}
             \end{bmatrix},\;i,j\in\mathbb{Z}_{\geq0},\quad\quad\mbox{ for all }\;n=0,1,2,\ldots,
$$
hence we obtain for every $i,j\in\mathbb{Z}_{\geq0}$ the expression
$$\breve{\Lambda}_{ji}(t)=(e^{t\breve{\mathcal{L}}})_{ji}=\sum_{n=0}^{\infty}\frac{t^n}{n!}\breve{\mathcal{L}}^{n}_{ji}
=\sum_{n=0}^{\infty}\frac{t^n}{n!}\begin{bmatrix}
\hat{\mathcal{L}}_{ji}^{n} & t^n\hat{\mathcal{L}}_{j,-i-1}^{n} \\
\hat{\mathcal{L}}_{-j-1,i}^{n} & \hat{\mathcal{L}}_{-j-1,-i-1}^{n}
             \end{bmatrix}=
\begin{bmatrix}
               \Lambda_{ji}(t) & \Lambda_{j,-i-1}(t) \\
               \Lambda_{-j-1,i}(t) & \Lambda_{-j-1,-i-1}(t)
             \end{bmatrix}.
$$

\end{proof}
\end{prop}

Note that we can evaluate $\breve{\Lambda}_{ji}(t)$ by \eqref{KMFqmc} and then extract the block $\Lambda_{ji}(t)$ as in \eqref{PHIbrevevshat}. Further, for a density operator $\rho$ we have
$$
p_{ji;\rho}(n)=\mathrm{Tr}\left(\Lambda_{ji}(t)\rho\right)=
\mathrm{Tr}\left(\begin{bmatrix}
\Lambda_{ji}(t) & 0 \\
 0 & 0
 \end{bmatrix}
\begin{bmatrix}
\rho  \\
0
\end{bmatrix}\right)
=\mathrm{Tr}\left(\begin{bmatrix}
I_{d^2} & 0 \\
0 & 0
\end{bmatrix}\breve{\Lambda}_{ji}(t)
\begin{bmatrix}
I_{d^2} & 0 \\
 0 & 0
 \end{bmatrix}
\begin{bmatrix}
\rho  \\
0
\end{bmatrix}\right).
$$
However, we would like to obtain the probability above avoiding the evaluation of $\breve{\Lambda}_{ji}(t).$ This can be done via a generalization of the Karlin-McGregor formula on $\mathbb{Z}_{\geq0}$. We proceed as follows: first, write the decomposition
$$dW(x)=\begin{bmatrix}
                              dW_{11}(x) & dW_{12}(x) \\dW_{21}(x) & dW_{22}(x)
                              \end{bmatrix},$$
where $dW_{21}(x)=dW_{12}^*(x)$, since $dW(x)$ is positive definite. Then one has for $i,j\in\mathbb{Z}_{\geq0},$
\begin{eqnarray}
\nonumber\breve{\Lambda}_{ji}(t)&=&\breve{\Pi}_j\int_\mathbb{R} e^{-xt}\mathcal{Q}_{j}^*(x)dW(x)\mathcal{Q}_{i}(x) \\
\nonumber   &\stackrel{\eqref{blobloQn}}{=}& \begin{bmatrix}\Pi_j & 0_{d^2} \\0_{d^2} & \Pi_{-j-1}\end{bmatrix}
\int_\mathbb{R} e^{-xt}\begin{bmatrix}
          Q_j^1(x) & Q_{-j-1}^1(x) \\
          Q_j^2(x) & Q_{-j-1}^2(x)
        \end{bmatrix}^*
\begin{bmatrix}
  dW_{11}(x) & dW_{12}(x) \\
  dW_{12}^*(x) & dW_{22}(x)
\end{bmatrix}
\begin{bmatrix}
          Q_i^1(x) & Q_{-i-1}^1(x) \\
          Q_i^2(x) & Q_{-i-1}^2(x)
        \end{bmatrix}
 \\
\nonumber   &=& \sum_{\alpha,\beta=1}^{2}
\begin{bmatrix}
\Pi_j\int_\mathbb{R}  e^{-xt}Q_j^{\alpha *}(x)dW_{\alpha\beta}(x)Q_i^{\beta}(x) & \Pi_j\int_\mathbb{R}  e^{-xt}Q_j^{\alpha *}(x)dW_{\alpha\beta}(x)Q_{-i-1}^{\beta }(x) \\
\Pi_{-j-1}\int_\mathbb{R}  e^{-xt}Q_{-j-1}^{\alpha *}(x)dW_{\alpha\beta}(x)Q_i^{\beta }(x) & \Pi_{-j-1}\int_\mathbb{R}  e^{-xt}Q_{-j-1}^{\alpha * }(x)dW_{\alpha\beta}(x)Q_{-i-1}^{\beta}(x)
\end{bmatrix} .
\end{eqnarray}
Joining equation above and Proposition \ref{Phibvsn}, we obtain the Karlin-McGregor formula for a CTOQW on $\mathbb{Z}$, given by
\begin{equation}\label{brevevsn}
\Lambda_{ji}(t)=\sum_{\alpha,\beta=1}^{2}\Pi_j\int_\mathbb{R} e^{-xt}Q_j^{\alpha *}(x)dW_{\alpha\beta}(x)Q_i^{\beta}(x),\;\;\mbox{for any}\; i,j\in \mathbb{Z},\;n\geq 0.
\end{equation}
Conversely, if there exist weight matrices $dW_{11}(x),dW_{12}(x),dW_{22}(x)$ such that $\Lambda_{ji}(t)$ is of the form \eqref{brevevsn}, then $\breve{\Lambda}_{ji}(t)$ is of the form
$$
\breve{\Phi}_{ji}^{(n)}=\breve{\Pi}_j\int_\mathbb{R} e^{-xt}\mathcal{Q}_{j}^*(x)dW(x)\mathcal{Q}_{i}(x).
$$
The weight matrix
\begin{equation*}\label{matrixspecphi}
W(x)=\begin{bmatrix}
            W_{11}(x) & W_{12}(x) \\
            W_{12}^*(x) & W_{22}(x)
          \end{bmatrix},
\end{equation*}
is called the spectral block matrix of $\mathcal{L}.$

\begin{remark}\label{remarkLIMstieltjes}
Extending Theorem \ref{CorRecurrence} to the CTOQW on $\mathbb{Z},$ we observe that, since $Q_0^1=Q_{-1}^2=I_d$ and $Q_0^2=Q_{-1}^1=0_d$, the following limits hold
$$
\int_{0}^{\infty}p_{00;\rho}(t)dt=\lim_{z\uparrow 0}\;\mathrm{Tr}\left[\Pi_0B(z;W_{11})vec(\rho)\right],
$$
where $B(z;W)$ is the Stieltjes transform of the weight matrix $W$. Analogously,
$$
\int_{0}^{\infty}p_{-1,-1;\rho}(t)dt=\lim_{z\uparrow 0}\;\mathrm{Tr}\left[\Pi_{-1}B(z;W_{22})vec(\rho)\right].
$$
\end{remark}

\medskip

Let us write the matrix $\hat{\mathcal{L}}$ in the form
\begin{equation*}\label{decmaismenos}
\breve{\mathcal{L}}=\begin{bmatrix}
             \hat{\mathcal{L}}^- & C \\
             A & \hat{\mathcal{L}}^+
           \end{bmatrix},\;\;
C=\begin{bmatrix}
    \vdots & \vdots & \vdots &  \\
    0 & 0 & 0 & \cdots \\
    \lceil C_0\rceil  & 0 & 0 & \cdots
  \end{bmatrix},\;\;
A=\begin{bmatrix}
    \cdots & 0 & 0 & \lceil A_{-1}\rceil  \\
    \cdots & 0 & 0 & 0 \\
    \cdots & 0 & 0 & 0 \\
     & \vdots & \vdots & \vdots
  \end{bmatrix},
\end{equation*}
$$
\breve{\mathcal{L}}^+=
\begin{bmatrix}
G^\alpha_{0}+\lceil B_{0}\rceil  & \lceil C_1\rceil  &  & & \\
  \lceil A_0\rceil &G^\alpha_{1}+\lceil B_{1}\rceil  & \lceil C_2\rceil    &&  \\
  &\lceil A_1\rceil &G^\alpha_{2}+\lceil B_{2}\rceil & \lceil C_3\rceil     & \\
   &  & \ddots &\ddots &\ddots
\end{bmatrix},$$
$$\breve{\mathcal{L}}^-=
\begin{bmatrix}
G^\alpha_{-1}+\lceil B_{-1}\rceil  & \lceil A_{-2}\rceil  &  & & \\
  \lceil C_{-1}\rceil &G^\alpha_{-2}+\lceil B_{-2}\rceil  & \lceil A_{-3}\rceil    &&  \\
  &\lceil C_{-2}\rceil &G^\alpha_{-3}+\lceil B_{-3}\rceil & \lceil A_3\rceil     & \\
   &  & \ddots &\ddots &\ddots
\end{bmatrix}
$$
Our goal now is to write the Stieltjes transforms associated with the weight matrices $W_{\alpha\beta},\alpha,\beta=1,2,$ in terms of the Stieltjes transforms associated with $W_{\pm}$, the weight matrices associated with $\breve{\mathcal{L}}^{\pm}$.

We introduce the generating function of $\hat{\mathcal{L}}$
$$\Phi(s):=\sum_{n=0}^{\infty}s^n\hat{\mathcal{L}}^n$$
to obtain an explicit form for the Laplace Transform of $\Lambda(t)$ on the following way:
$$\widehat{\Lambda}_{ji}(t)= \int_{0}^{\infty}e^{-xt}\Lambda_{ji}(x)dx=\sum_{n=0}^{\infty}\int_{0}^{\infty}e^{-xt}\frac{x^n}{n!}\hat{\mathcal{L}}_{ji}^ndx= \sum_{n=0}^{\infty}\frac{\widehat{t^n}}{n!}\hat{\mathcal{L}}_{ji}^n=\sum_{n=0}^{\infty}\frac{\hat{\mathcal{L}}_{ji}^n}{t^{n+1}}=\frac{\Phi_{ji}(t^{-1})}{t}.
$$

Using equations (48), (49), (50) and (51) of \cite{dILL}, applied to $\Phi_{ji}(s^{-1})=s\hat{\Lambda}_{ji}(s)$, we obtain
\begin{eqnarray}
 \widehat{\Lambda}_{00}(z)  &=&\widehat{\Lambda}_{00}^+(z)(I-\lceil A_{-1}\rceil \widehat{\Lambda}_{-1,-1}^-(z)\lceil C_0\rceil \widehat{\Lambda}_{00}^+(z))^{-1}.
 \label{PP1}\\
 \widehat{\Lambda}_{-1,-1}(z)  &=&\widehat{\Lambda}^-_{-1,-1}(z)(I-\lceil C_0\rceil \widehat{\Lambda}^+_{00}(z)\lceil A_{-1}\rceil \widehat{\Lambda}_{-1,-1}^-(z))^{-1}.
 \label{PP2}\\
 \widehat{\Lambda}_{0,-1}(z)  &=& z^{-1}\widehat{\Lambda}_{00}^+(z)(I-\lceil A_{-1}\rceil \widehat{\Lambda}_{-1,-1}^-(z)\lceil C_0\rceil \widehat{\Lambda}_{00}^+(z))^{-1}\lceil A_{-1}\rceil \widehat{\Lambda}_{-1,-1}^-(z).
 \label{PP3}\\
 \widehat{\Lambda}_{-1,0}(z)  &=&  z^{-1}\widehat{\Lambda}^-_{-1,-1}(z)(I-\lceil C_0\rceil \widehat{\Lambda}^+_{00}(z)\lceil A_{-1}\rceil \widehat{\Lambda}_{-1,-1}(z))^{-1}\lceil C_0\rceil \widehat{\Lambda}_{00}^+(z).
\label{PP4}
\end{eqnarray}

We notice that the block matrices of both $\breve{\mathcal{L}}^+$ and $\breve{\mathcal{L}}^-$ satisfy the conditions of equation \eqref{condiequiv}, thus there are positive weight matrices $W_\pm$ associated with $\breve{\mathcal{L}}^\pm$ for which the associated polynomials are orthogonal. Then, we can write
$$
\Pi_0^+:=\int_\mathbb{R}dW_+\;\;\;\mbox{ and }\;\;\;\Pi_{-1}^-:=\int_\mathbb{R}dW_-\;\;.
$$

The Laplace Transform of $\Lambda_{ji}(t)$ can be associated to the Stieltjes transform using that
$$\widehat{\Lambda}_{ji}(s) = \int_{0}^{\infty}e^{-ts}\Lambda_{ji}(t)dt=\int_{0}^{\infty}e^{-ts}\left(\Pi_j\int_\mathbb{R}e^{-xt}Q_j^*(x)dW(x)Q_i(x)dt\right)= \Pi_j\int_\mathbb{R} \frac{Q_j^*(x)dW(x)Q_i(x)}{s+x},$$
$s>0,$ that is,
$$
\widehat{\Lambda}_{ji}(-s)=\Pi_j\int_\mathbb{R} \frac{Q_j^*(x)dW(x)Q_i(x)}{x-s},\;s<0,
$$
thereby we recall that $Q_0^1=Q_{-1}^2=I_{d^2},$ $Q_0^2=Q_{-1}^1=0_{d^2}$ in order to obtain the relations
$$
\begin{array}{lll}
  B(z;W_{11})=\Pi_0^{-1}\widehat{\Lambda}_{00}(-z),&B(z;W_{22})=\Pi_{-1}^{-1}\widehat{\Lambda}_{-1,-1}(-z),&B(z^{-1};
  W_{12})=\Pi_{-1}^{-1}\widehat{\Lambda}_{0,-1}(-z), \\
  B(z;W_{21})=\Pi_{-1}^{-1}\widehat{\Lambda}_{-1,0}(-z),&B(z;W_+)=(\Pi^{+}_0)^{-1}\widehat{\Lambda}_{00}^+(-z),&
  B(z^{-1};W_-)=(\Pi^{-}_{-1})^{-1}\widehat{\Lambda}_{-1,-1}^-(-z).
\end{array}
$$
Joining with the identities \eqref{PP1},\eqref{PP2},\eqref{PP3},\eqref{PP4},  the new Stieltjes transform identities are obtained:
\begin{equation}\label{newstieltjes}
\begin{split}
\Pi_{0} B(z;W_{11}) &= \Pi_{0}^+B(z;W_+)(I-\lceil A_{-1}\rceil \Pi_{-1}^-B(z;W_-)\lceil C_0\rceil \Pi_{0}^+B(z;W_+))^{-1}, \\
\Pi_{-1} B(z;W_{22})&=\Pi_{-1}^-B(z;W_-)(I-\lceil C_0\rceil \Pi_{0}^+B(z;W_+)\lceil A_{-1}\rceil \Pi_{-1}^-B(z;W_-))^{-1},  \\
 \Pi_{0} B(z;W_{12}) &= \Pi_{0}^+B(z;W_+)(I-\lceil A_{-1}\rceil \Pi_{-1}^-B(z;W_-)\lceil C_0\rceil \Pi_{0}^+B(z;W_+))^{-1}\lceil A_{-1}\rceil \Pi_{-1}^-B(z;W_-), \\
 \Pi_{-1} B(z;W_{21}) &=\Pi_{-1}^-B(z;W_-)(I-\lceil C_0\rceil \Pi_{0}^+B(z;W_+)\lceil A_{-1}\rceil \Pi_{-1}^-B(z;W_-))^{-1}\lceil C_0\rceil \Pi_{0}^+B(z;W_+).
\end{split}
\end{equation}
Sometimes the operators $\Pi_i^+$ and $\Pi_i^-$ are equal to the identity operator. In this case, \eqref{newstieltjes} are reduced to
\beq\label{newstieltjesORTH}
\begin{split}
B(z;W_{11}) &= B(z;W_+)(I-\lceil A_{-1}\rceil B(z;W_-)\lceil C_0\rceil B(z;W_+))^{-1}, \\
B(z;W_{22}) &=B(z;W_-)(I-\lceil C_0\rceil B(z;W_+)\lceil A_{-1}\rceil B(z;W_-))^{-1},  \\
B(z;W_{12}) &=  B(z;W_+)(I-\lceil A_{-1}\rceil B(z;W_-)\lceil C_0\rceil B(z;W_+))^{-1}\lceil A_{-1}\rceil B(z;W_-), \\
B(z;W_{21}) &= B(z;W_-)(I-\lceil C_0\rceil B(z;W_+)\lceil A_{-1}\rceil B(z;W_-))^{-1}\lceil C_0\rceil B(z;W_+).
\end{split}
\eeq

Equations \eqref{newstieltjes} and \eqref{newstieltjesORTH} allow us to obtain the Stieltjes transform of the CTOQW with $V=\mathbb{Z}$ when we know the Stieltjes transform associated to the walks on $\mathbb{Z}_\geq 0$ and $\mathbb{Z}_\leq 0.$ Since we are interested in the recurrence and transience of the CTOQWs, those equations are enough to obtain this information as it will be seen on the next section.

\begin{remark}
A sufficient condition for $\Pi_i^+=\Pi_i^-=I$ is to have $A_n=C_{n+1}^*$ and $B_n=B_n^*$ for every $n\in\mathbb{Z},$ since we will always have $G_n=G_n^*$ for all $n\in\mathbb{Z}$ in this case, and therefore we can take $R_i=I$ for all $i\in \mathbb{Z}$ (see Equation \eqref{condiequiv}). On the other hand, those conditions are not necessary, since we can find examples with $R_n$ being any unitary matrices for each $n$.
\end{remark}

\medskip

Most of our examples consider $R_i^i=0$ for all $i\in V.$ In this case the Hamiltonian part does not contribute to the probabilities, as it will be seen as a consequence of the following Proposition. Moreover, this Proposition gives equivalence to a condition that the diagonal of the matrix representation of the generator has negative-semidefinite matrices.

\begin{prop}\label{propHh}
Let us consider a tridiagonal  CTOQW in $Z_{\geq 0}$ (or a finite $V$) satisfying the conditions of Equation \eqref{EqDette}. Then $G_n^\alpha+\lceil B_n\rceil\leq 0$ if and only if
\begin{equation}\label{startH}
-H_n\otimes I+I\otimes \overline{H_n}=
\begin{bmatrix}
  h_{11}^{(n)} & \ldots & h_{1,d^2}^{(n)} \\
  \vdots & \ddots & \vdots \\
  h_{d^2,1}^{(n)} & \ldots & h_{d^2,d^2}^{(n)}
\end{bmatrix},\quad h_{kk}^{(n)}=-b_{kk}^{(n)},\quad h_{jk}^{(n)}=-i\left(s_{jk}^{(n)}-a_{jk}^{(n)}-ib_{jk}^{(n)}\right),\;\; \forall\; j,k,
\end{equation}
where
$$
\lceil B_n\rceil=\begin{bmatrix}
  a_{11}^{(n)} & \ldots & a_{1,d^2}^{(n)} \\
  \vdots & \ddots & \vdots \\
  a_{d^2,1}^{(n)} & \ldots & a_{d^2,d^2}^{(n)}
\end{bmatrix}+
i\begin{bmatrix}
  b_{11}^{(n)} & \ldots & b_{1,d^2}^{(n)} \\
  \vdots & \ddots & \vdots \\
  b_{d^2,1}^{(n)} & \ldots & b_{d^2,d^2}^{(n)}
\end{bmatrix},\quad a_{jk},b_{jk}\in\mathbb{R}\;\forall\; j,k
$$
and
$$
S_n\otimes I+I\otimes \overline{S_n}=
\begin{bmatrix}
  s_{11}^{(n)} & \ldots & s_{1,d^2}^{(n)} \\
  \vdots & \ddots & \vdots \\
  s_{d^2,1}^{(n)} & \ldots & s_{d^2,d^2}^{(n)}
\end{bmatrix}\;\forall\; j,k,\quad S_n:=\frac{1}{2}\left(A_n^*A_n+B_n^*B_n+C_n^*C_n\right).
$$

\begin{proof}Let us suppose that $T_n:=G_n^\alpha+\lceil B_n\rceil$ for every $n\geq 0$ satisfies the conditions of Equation \eqref{EqDette}.

Firstly we suppose that $T_n\leq 0,$ thus there exists an orthonormal basis $\{v_1,\ldots,v_{d^2}\}$ of $\mathbb{C}^{d^2}$ constituted by eigenvectors of $T_n$ with $T_nv_k=t_kv_k,\;k=1,\ldots,d^2.$

Denote $S_n=\frac{1}{2}\left(A_n^*A_n+B_n^*B_n+C_n^*C_n\right)$ to obtain
\begin{equation*}
\begin{split}
t_k\delta_{jk}= t_k\langle v_j,v_k\rangle=& \langle v_j,(G_n^\alpha+\lceil B_n\rceil) v_k\rangle=\langle v_k,\left(-iH_n\otimes I+iI\otimes \overline{H_n}-S_n\otimes I-I\otimes \overline{S_n}+\lceil B_n\rceil\right) v_k\rangle\\
=& i\underbrace{\langle v_j,\left(-H_n\otimes I+I\otimes \overline{H_n}\right) v_k\rangle}_{F_{1,j,k}}-\underbrace{\langle v_j,\left(S_n\otimes I+I\otimes \overline{S_n}\right) v_k\rangle}_{F_{2,j,k}}+\underbrace{\langle v_j,\lceil B_n\rceil v_k\rangle}_{F_{3,j,k}}.
\end{split}
\end{equation*}

We have $F_{1,k,k},F_{2,k,k}\in \mathbb{R},$ thus $\langle v_k,\left(-H_n\otimes I+I\otimes \overline{H_n}\right) v_k\rangle=-Im(\langle v_k,\lceil B_n\rceil v_k\rangle),$ thereby the entries of the diagonal of $-H_n\otimes I+I\otimes \overline{H_n}$ coincide with the entries of the imaginary part of the diagonal of $-\lceil B_n\rceil.$

For $j\neq k,$ we have
$$
i\langle v_j,\left(-H_n\otimes I+I\otimes \overline{H_n}\right) v_k\rangle=\langle v_j,\left(S\otimes I+I\otimes \overline{S}\right) v_k\rangle-\langle v_j,\lceil B_n\rceil v_k\rangle,
$$
thus, denoting by $[X]_{jk}$ the $(j,k)$-th entry of a matrix $X$ on the basis $(v_k)_k,$ we obtain the identity
$$
i[-H_n\otimes I+I\otimes \overline{H_n}]_{jk}=[S\otimes I+I\otimes \overline{S}-\lceil B_n\rceil]_{j,k},\quad j\neq k,
$$
completing the first part of the proof.

On the other hand, we suppose that there exists an orthonormal basis such that equation \eqref{startH} is valid. In this case we have
$$
[T_n]_{kk}=ih_{kk}^{(n)}-s_{kk}^{(n)}+a_{kk}^{(n)}+ib_{kk}^{(n)}=-s_{kk}^{(n)}+a_{kk}^{(n)}<0,\quad \forall k,
$$
and
$$
[T_n]_{jk}=ih_{jk}^{(n)}-s_{jk}^{(n)}+a_{jk}^{(n)}+ib_{jk}^{(n)}=i\left(-i\left(s_{jk}^{(n)}-a_{jk}^{(n)}-ib_{jk}^{(n)}\right)\right)-s_{jk}^{(n)}+a_{jk}^{(n)}+ib_{jk}^{(n)}=0\;\forall j\neq k.
$$

This shows that $[T_n]$ is diagonal with respect to this orthonormal basis and its entries are all real, thus it is hermitian.
\end{proof}
\end{prop}

\begin{corollary}Let us consider a tridiagonal  CTOQW in $Z_{\geq 0}$ (or a finite $V$) with a positive matrix weight associated to this CTOQW. Then $H_n=h_nI$ for some $h_n\in R$ if and only if $B_n$ is hermitian. In this case, $H_n$ does not contribute to the probability of the walk.
\begin{proof}
We suppose that there exists a positive matrix weight associated to the CTOQW, thus Equation \eqref{startH} is valid. We have that $B_n$ is hermitian if and only if
$$
a_{jk}^{(n)}+ib_{jk}^{(n)}=a_{kj}^{(n)}-ib_{kj}^{(n)}\;\forall j,k.
$$
The matrix $-H_n\otimes I+I\otimes \overline{H_n}$ is a multiple of the identity if and only if $h_{jk}^{(n)}=0\;\forall j\neq k$ and $h_{kk}^{(n)}=h\;\forall k$ for some $h\in\mathbb{R},$ where the second statement is valid by Proposition \ref{propHh}. Moreover,
$$
h_{jk}^{(n)}=0\Leftrightarrow s_{jk}^{(n)}=a_{jk}^{(n)}+ib_{jk}^{(n)},
$$
which is equivalent to have $\lceil B_n\rceil$ to be hermitian, since $S$ is hermitian.
\end{proof}
\end{corollary}

\section{Examples}

\subsection{Diagonal and simultaneously diagonalizable transitions}\label{exDiag}

{\color{black}First, we will consider a homogeneous CTOQW on the $N+1$ nodes indexed as $V=\{0,1,\dots, N\}$, where we add two absorbing barriers ($\ket{-1},\;\ket{N+1}$) on the extreme nodes}, $R_i^i=0$ for each site, and the generator $\mathcal{L}$ is given by
\begin{equation*}
\hat{\mathcal{L}}=\begin{bmatrix} G^\alpha & \lceil C\rceil  &  & & \\ \lceil A\rceil  & G^\alpha& \lceil C\rceil   & & \\  & \lceil A\rceil  & G^\alpha & \lceil C\rceil   & \\ & \ddots & \ddots & \ddots &  \\ &  & \lceil A\rceil  & G^\alpha & \lceil C\rceil \\  & &  & \lceil A\rceil  & G^\alpha\end{bmatrix},\;\;A=\begin{bmatrix}a_1 & 0 \\0 & a_2\end{bmatrix},\;\;C=\begin{bmatrix}c_1 & 0 \\0 & c_2\end{bmatrix},\;\;a_1,a_2,c_2,c_2>0,
\end{equation*}

$$
G^\alpha=-\mathrm{diag}\left(a_1^2+c_1^2,\frac{a_1^2+c_1^2+a_2^2+c_2^2}{2},\frac{a_1^2+c_1^2+a_2^2+c_2^2,a_2^2+c_2^2}{2},a_2^2+c_2^2\right).
$$
The classical symmetrization
$$\mathcal{R}=\mbox{diag}(R_0,R_1,\dots,R_{N}),\;\;\;R_i=K^{\frac{i-1}{2}},\;\;\;i=1,\dots,N,\quad R_0=I_4,$$
where $K=\lceil \sqrt{AC}\rceil =\mathrm{diag}\left(a_1c_1,\;\sqrt{a_1c_1a_2c_2},\;\sqrt{a_1c_1a_2c_2},\;a_2c_2\right),$ gives
$$J=\mathcal{R}\hat{\mathcal{L}}\mathcal{R}^{-1}=\begin{bmatrix} G^\alpha & K &  & & \\ K & G^\alpha & K  & & \\  & K & G^\alpha & K  & \\ & \ddots & \ddots & \ddots &  \\ &  & K & G^\alpha & K\\  & &  & K & G^\alpha\end{bmatrix}.
$$

The matrix-valued polynomials $\{Q_n\}_{n\geq0}$ are recursively defined by
\begin{align*}
Q_0(x)&=1,\quad Q_{-1}(x)=0,\\
-xQ_0(x)&=Q_0(x)G^\alpha+Q_1(x)K,\\
-xQ_i(x)&=Q_{i+1}(x)K+Q_i(x)G^\alpha+Q_{i-1}(x)K,\quad i=1,\dots,N-1,
\end{align*}
which can be identified with the Chebyshev polynomials of the second kind $\{U_n\}_{n\geq0}$. Indeed, we have
$$
Q_n(x)=U_n\left(\frac{(-x-G^\alpha)K^{-1}}{2}\right),\quad n\geq 0.
$$

Now, if we define
\begin{equation*}
R_{N+1}(x):=Q_N(x)(-x-G^\alpha)-Q_{N-1}(x)K,
\end{equation*}
we have that the zeros of  $\det(R_{N+1}(x))$ coincide with the eigenvalues of $-J$. A simple calculation shows that
$$R_{N+1}(x)=U_{N+1}\left(\frac{(-x-G^\alpha)K^{-1}}{2}\right)K.$$
We would like to solve the equation det$(R_{N+1}(x))=0$. Recalling the representation
$$U_n\left(\frac{z}{2}\right)=\prod_{j=1}^n\left(z-2\cos\left(\frac{j\pi}{n+1}\right)\right),$$
we obtain, for the matrix-valued case at hand,
$$
\mbox{det}(R_{N+1}(x))=\mbox{det}\left(U_{N+1}\left(\frac{(-x-G^\alpha)K^{-1}}{2}\right)K\right)=\mbox{det}\left[\prod_{j=1}^{N+1}\left((-xI_4-G^\alpha)K^{-1}-2\cos\left(\frac{j\pi}{N+2}\right)\right)K\right],
$$
thus
$$
\mbox{det}(R_{N+1}(x))=k_1k_2^2k_4\prod_{j=1}^{N+1}\prod_{m=1}^{4}\left[\frac{(-x-g_m)}{k_m}-2\cos\left(\frac{j\pi}{N+2}\right)\right],
$$
where we have put $G=-\mathrm{diag}(g_1,g_2,g_3,g_4)$ and $K=-\mathrm{diag}(k_1,k_2,k_3,k_4).$ Since $g_2=g_3$ and $k_2=k_3,$ $\mbox{det}(R_{N+1}(x))$ is a polynomial of degree $4(N+1)$ having $3(N+1)$ distinct roots, which are of the form
\begin{equation*}
\begin{split}
x_{j}=& -g_1-2k_1\cos\left(\pi\frac{j+1}{N+2}\right)=a_1^2+c_1^2-2a_1c_1\cos\left(\pi\frac{j+1}{N+2}\right),\\
y_{j}=& -g_2-2k_2\cos\left(\pi\frac{j+1}{N+2}\right)=\sqrt{a_1c_1a_2c_2}-(a_1^2+c_1^2+a_2^2+c_2^2)\cos\left(\pi\frac{j+1}{N+2}\right),\\
z_{j}=& -g_4-2k_4\cos\left(\pi\frac{j+1}{N+2}\right)=a_2^2+c_2^2-2a_2c_2\cos\left(\pi\frac{j+1}{N+2}\right),\;\;\;j=0,\dots,N,
\end{split}
\end{equation*}
each $y_j$ being of multiplicity 2. There can be cases of eigenvalues with a greater multiplicity, which happens when the collection of zeros $x_N,y_N$ and $z_N$ overlap, so the multiplicity changes accordingly.

Let us compute the weight matrixes on the zeros above. We have
\beq\label{doublef}
W_j=g_j'(\lambda_j),\;\;\;g_j(\lambda):=-(\lambda_j-\lambda)^2(-J-\lambda I)_{00}^{-1},\; \lambda_j=x_j,y_j,z_j,\; j=0,\dots,N,
\eeq
an expression which can be deduced from (see \cite{grunb})
$$
(-J-\lambda I)_{ij}^{-1}=\sum_{k=0}^N \frac{P_i^*(\lambda_k)W_k P_j(\lambda_k)}{\lambda_k-\lambda},
$$
and noting that this corresponds to the Laurent sum of the operator on the left-hand side except for the sign change $\lambda_k-\lambda=-(\lambda-\lambda_k)$. With formula \eqref{doublef}, a calculation shows that for every $N$ we have a corresponding set of multiples of the matrices given by
$$
W_{K;1}=\begin{bmatrix}1 & 0 & 0 & 0\\0 & 0 & 0 & 0\\0 & 0 & 0 & 0\\0 & 0 & 0 & 0 \end{bmatrix},\;
W_{K;2}=\begin{bmatrix}0 & 0 & 0 & 0\\0 & 0 & 0 & 0\\0 & 0 & 0 & 0\\0 & 0 & 0 & 1 \end{bmatrix},\;
W_{K;3}=\begin{bmatrix}0 & 0 & 0 & 0\\0 & 1 & 0 & 0\\0 & 0 & 1 & 0\\0 & 0 & 0 & 0 \end{bmatrix}.
$$
More precisely, we have a collection of $3(N+1)$ roots with weights
$$\psi(x_{j})=\frac{2}{N+2}\sin^2\left(\pi\frac{j+1}{N+2}\right)W_{K;1},\;\;\;j=0,\dots,N,$$
$$\psi(y_{j})=\frac{2}{N+2}\sin^2\left(\pi\frac{j+1}{N+2}\right)W_{K;2},\;\;\;j=0,\dots,N.$$
$$\psi(z_{j})=\frac{2}{N+2}\sin^2\left(\pi\frac{j+1}{N+2}\right)W_{K;2},\;\;\;j=0,\dots,N.$$

For a specific instance of the above take $N=2$ (3 sites), so we have 9 roots, with weights
$$\frac{1}{4}W_{K;1},\;\;\;\frac{1}{4}W_{K;2},\;\;\;\frac{1}{4}W_{K;3}$$
associated with zeros $\;a_1^2+c_1^2-2a_1c_1,\;\sqrt{a_1c_1a_2c_2}-(a_1^2+c_1^2+a_2^2+c_2^2)$ and $\;a_2^2+c_2^2-2a_2c_2\;$ respectively; weights
$$\frac{1}{2}W_{K;1},\;\;\;\frac{1}{2}W_{K;2},\;\;\;\frac{1}{2}W_{K;3}$$
associated with zeros $\;a_1^2+c_1^2-\sqrt{2}a_1c_1,\;\sqrt{a_1c_1a_2c_2}-\sqrt{2}(a_1^2+c_1^2+a_2^2+c_2^2)/2$ and $\;a_2^2+c_2^2-\sqrt{2}a_2c_2\;$ respectively; and weights
$$\frac{1}{4}W_{K;1},\;\;\;\frac{1}{4}W_{K;2},\;\;\;\frac{1}{4}W_{K;3}$$
associated with zeros $\;a_1^2+c_1^2,\;\sqrt{a_1c_1a_2c_2}$ and $\;a_2^2+c_2^2\;$ respectively.

\medskip

{\bf Now, let us consider the walk on the half-line.}

\medskip

{\color{black}
We will consider a CTOQW whose set of vertices is $V=\{0,1,2,\ldots\}$ and the walker can jump to its nearest neighbor, however, there is an absorbing barrier ($\ket{-1}$). Thereofore, this walk can be interpreted as a BDP in which the population may become extinct.} The matrix
$$\hat{\mathcal{L}}=
\begin{bmatrix}
  G^\alpha_0 & \lceil C\rceil  &  &  &  \\
  \lceil A\rceil  & G^\alpha & \lceil C\rceil  &  &  \\
  &\lceil A\rceil  & G^\alpha & \lceil C\rceil  &   \\
   &  & \ddots & \ddots & \ddots
\end{bmatrix},\;
\begin{array}{ll}
 G^\alpha= & -\frac{1}{2}\left((A^*A+C^*C)\otimes I_2+I_2\otimes(A^*A+C^*C)\right) \\
 &\\
  G^\alpha_0= & -\frac{1}{2}\left((A^*A)\otimes I_2+I_2\otimes(A^*A)\right),
\end{array}
$$
is a valid generator of a CTOQW. Also,
$$G^\alpha=-\begin{bmatrix}a_1^2+c_1^2& 0 & 0 & 0 \\0& \frac{a_1^2+c_1^2+a_2^2+c_2^2}{2} & 0 & 0 \\0& 0 & \frac{a_1^2+c_1^2+a_2^2+c_2^2}{2} & 0 \\0& 0 & 0 & a_2^2+c_2^2 \end{bmatrix},
$$
$$G^\alpha_0=-\begin{bmatrix}a_1^2& 0 & 0 & 0 \\0& \frac{a_1^2+a_2^2}{2} & 0 & 0 \\0& 0 & \frac{a_1^2+a_2^2}{2} & 0 \\0& 0 & 0 & a_2^2\end{bmatrix}.
$$

{\color{black} The CTOQWs of these first examples are entirely described by diagonal matrices; therefore, the parameter $b$ in the density matrix $\rho$ has no influence on these random walks with the specific transitions $A$ and $C$.}

If we take $K:=\lceil (AC)\rceil ^{1/2}$ then we obtain the symmetrization
$$
J=\mathcal{R}(-\hat{\mathcal{L}})\mathcal{R}^{-1}=\begin{bmatrix} -G^\alpha_0 & K &  & & \\ K & -G^\alpha & K  & & \\  & K & -G^\alpha & K  & \\ &  & \ddots & \ddots &\ddots\end{bmatrix},
$$
where $K$ is positive definite,
$$\mathcal{R}=\mbox{diag}(R_0,R_1,\dots,R_{N}),\;\;\;R_i=\lceil A^{-1}C\rceil ^{i-1},\;\;\;i=1,2,3,\dots,N,\quad R_0=I_4.$$

Let us obtain the weight matrix associated to $\tilde{J},$
$$\tilde{J}:=\begin{bmatrix} -G^\alpha & K &  & & \\ K & -G^\alpha & K  & & \\  & K & -G^\alpha & K  & \\ &  & \ddots & \ddots &\ddots\end{bmatrix},$$
using the results of A.J. Dur\'an (\cite{duran-ratio}).

Since $G^\alpha$ and $K$ commute it is easy to see that the matrix $H_{A,B}(x)$ given by \cite{duran-ratio} is
\begin{equation}\nonumber
\begin{split}
 H(x)=   & (xI+G^\alpha)^2K^{-2}-4I_4=(xI+G^\alpha)^2\lceil AC\rceil ^{-1}-4I_4 \\
   = & \begin{bmatrix}\frac{(x-a_1^2-c_1^2)^2}{a_1^2c_1^2}-4& 0 & 0 & 0 \\0& \frac{(x-\frac{a_1^2+c_1^2+a_2^2+c_2^2}{2})^2}{a_1a_2c_1c_2}-4 & 0 & 0 \\0& 0 & \frac{(x-\frac{a_1^2+c_1^2+a_2^2+c_2^2}{2})^2}{a_1a_2c_1c_2}-4 & 0 \\0& 0 & 0 & \frac{(x-a_2^2-c_2^2)^2}{a_2^2c_2^2}-4 \end{bmatrix}.
\end{split}
\end{equation}

The associated weight matrix to $\tilde{J}$ is
\begin{equation}\label{exTilde}
d\tilde{\Sigma}(x)= \frac{1}{2\pi}K^{-1}\sqrt{\mathrm{diag}(h_1,h_2,h_3,h_4)}
=\frac{1}{2\pi}\begin{bmatrix}d_1(x)& 0 & 0 & 0 \\0&d_2(x) & 0 & 0 \\0& 0 & d_3(x) & 0 \\0& 0 & 0 & d_4(x)\end{bmatrix},
\end{equation}
where $h_j$ represents the $j$-th diagonal entry of the diagonal appearing on the representation of $H(x)$ and
\begin{equation}\nonumber
\begin{split}
d_1(x)= & \frac{\left[\sqrt{4a_1^2c_1^2-(x-a_1^2-c_1^2)^2}\right]_+}{a_1^2c_1^2},\;\; d_4(x)=\frac{\left[\sqrt{4a_2^2c_2^2-(x-a_2^2-c_2^2)^2}\right]_+}{a_2^2c_2^2}\\
d_2(x)=   & d_3(x)=\frac{\left[\sqrt{4a_1a_2c_1c_2-\left(x-\frac{a_1^2+c_1^2+a_2^2+c_2^2}{2}\right)^2}\right]_+}{2a_1a_2c_1c_2}.
\end{split}
\end{equation}
Here we are using the notation $[f(x)]_+ = f(x)$ if $f(x)\geq0$ and 0 otherwise.

We are interested on the transitions of the CTOQW, thus only $d_1(x)$ and $d_4(x)$ contribute for the calculus of the trace when we evaluate
$$\textmd{Tr}\left(\begin{bmatrix}d_1(x)& 0 & 0 & 0 \\0&d_2(x) & 0 & 0 \\0& 0 & d_3(x) & 0 \\0& 0 & 0 & d_4(x)\end{bmatrix}vec(\rho)\right),$$
thereby we will avoid the massive calculations using terms as $d_2(x)$ and $d_3(x)$ appearing on the sequel of this section.

The Stieltjes transform is
\begin{equation}\label{Bs1}
B(z,\tilde{\Sigma})= K^{-1}\sqrt{\mathrm{diag}(h_1,h_2,h_3,h_4)}
=\begin{bmatrix}w_1(z)& 0 & 0 & 0 \\0&w_2(z) & 0 & 0 \\0& 0 & w_3(z) & 0 \\0& 0 & 0 & w_4(z)\end{bmatrix},
\end{equation}
where $w_2(z)=w_3(z)$ is a function that does not vanish and
\begin{equation}\nonumber
\begin{split}
w_1(z)=&\frac{z-a_1^2-c_1^2-i\sqrt{4a_1^2c_1^2-(z-a_1^2-c_1^2)^2}}{2a_1^2c_1^2}, \\
w_4(z)=&\frac{z-a_2^2-c_2^2-i\sqrt{4a_2^2c_2^2-(z-a_2^2-c_2^2)^2}}{2a_2^2c_2^2}.
\end{split}
\end{equation}

Since the weight is obtained on the terms of \cite{duran-ratio}, we must have $\Pi_0=I_4,$ then we use equation (2.20) of \cite{detteC} to obtain the Stieltjes transform of the weight matrix associated to $J$:
\begin{equation*}\label{SDe}
B(z,\Sigma)= \left(B(z,\tilde{\Sigma})^{-1}+(G^\alpha_0-G^\alpha)\right)^{-1}=\begin{bmatrix}\sigma_1(z)& 0 & 0 & 0 \\0&\ast & 0 & 0 \\0& 0 & \ast & 0 \\0& 0 & 0 & \sigma_2(z)\end{bmatrix},
\end{equation*}
where
$$\sigma_j(z)=\frac{z-a_j^2+c_j^2+\sqrt{-4a_j^2c_j^2+(z+a_j^2+c_j^2)}}{2c_j^2z},\;\;j=1,2.$$

It is a simple calculation to verify that $\lim_{z\uparrow 0}\sigma_j(z)=\infty\Leftrightarrow a_j\leq c_j,$ thus, given a density operator $\rho=\begin{bmatrix}a & b \\b^* & 1-a\end{bmatrix},$ we have
$$
\lim_{z\uparrow 0}\textmd{Tr}\left[vec^{-1}\Pi_0\left(B(z,\Sigma)vec(\rho)\right)\right]=\lim_{z\uparrow 0}\left(\pi_1\sigma_1(z)a+\pi_2\sigma_2(z)(1-a)\right),
$$
where $\pi_1,\pi_2>0.$
Therefore, if $\{\ket{e_0},\ket{e_1}\}$ is the canonical basis of $\mathbb{C}^2$, then an application of Corollary \ref{CorRecurrence} shows that
\begin{itemize}
  \item $a_1\leq c_1$ and $a_2\leq c_2\;\Rightarrow$ vertex $\ket{0}$ is recurrent;
  \item $a_1\leq c_1$ and $a_2> c_2\;\Rightarrow$ vertex $\ket{0}$ is $\ket{e_1}\bra{e_1}$-transient and $\rho$-recurrent for $\rho\neq \ket{e_1}\bra{e_1};$
  \item $a_1> c_1$ and $a_2\leq c_2\;\Rightarrow$ vertex $\ket{0}$ is $\ket{e_0}\bra{e_0}$-transient and $\rho$-recurrent for $\rho\neq \ket{e_0}\bra{e_0};$
  \item $a_1> c_1$ and $a_2>c_2\;\Rightarrow$ vertex $\ket{0}$ is transient.
\end{itemize}

The Perron-Stieltjes inversion formula (Proposition 1.1 of \cite{Manuel}) gives
$$d\Sigma(x)=\frac{1}{\pi}
\begin{bmatrix}\left[\frac{\sqrt{4a_1^2c_1^2-(x-a_1^2-c_1^2)^2}}{2c_1^2x}\right]_+& 0 & 0 & 0 \\0&\ast & 0 & 0 \\0& 0 & \ast & 0 \\0& 0 & 0 & \left[\frac{\sqrt{4a_2^2c_2^2-(x-a_2^2-c_2^2)^2}}{2c_2^2x}\right]_+\end{bmatrix},
$$
thus an application of the Karlin-McGregor formula for CTOQWs gives for $\rho=\begin{bmatrix}a & b \\b^* & 1-a\end{bmatrix},$
$$
p_{00;\rho}(t)=a\int_0^\infty e^{-xt}\left[\frac{\sqrt{4a_1^2c_1^2-(x-a_1^2-c_1^2)^2}}{2c_1^2x}\right]_+dx+
(1-a)\int_0^\infty e^{-xt}\left[\frac{\sqrt{4a_2^2c_2^2-(x-a_2^2-c_2^2)^2}}{2c_2^2x}\right]_+dx.
$$

The particular case of $r:=a_1=c_1$ and $s:=a_2=c_2$ gives the weight matrix
$$d\Sigma(x)=\frac{1}{\pi}
\begin{bmatrix}\left[\frac{\sqrt{-x^2+4xr^2}}{2r^2x}\right]_+& 0 & 0 & 0 \\0&w_{r,s}(x) & 0 & 0 \\0& 0 & w_{r,s}(x) & 0 \\0& 0 & 0 & \left[\frac{\sqrt{-x^2+4xs^2}}{2s^2x}\right]_+
\end{bmatrix},
$$
where
$$
w_{r,s}(x)=\left[\frac{2\sqrt{((r+s)^2-x)(x-(r-s)^2)}}{2(r^2+s^2)x-(r^2-s^2)^2}\right]_++\left(\frac{(r+s)(r-s)}{r^2+s^2}\right)^2\delta_{x_0}(z)
,\;\;x_0=\frac{(r+s)^2(r-s)^2}{2(r^2+s^2)}.$$

{\bf Finally, we describe the associated walk on the integer line.}

\medskip

Let us consider the homogeneous CTOQW on $\mathbb{Z}.$ {\color{black}In this case, the quantum walker's dynamics are uniform across different positions on the integer lattice, and could be explored in various physical systems, such as trapped ions, superconducting circuits, or photonic systems.}

We take
$$R_{i}^{i+1}=A=\begin{bmatrix}a_1 & 0 \\0 & a_2\end{bmatrix},\quad R_{i}^{i-1}=C=\begin{bmatrix}c_1 & 0 \\0 & c_2\end{bmatrix},\;\forall i\in\mathbb{Z},\quad a_1,a_2,c_1,c_2>0.$$
In this case we have
$$G_i=-\begin{bmatrix}a_1^2+c_1^2& 0 & 0 & 0 \\0& \frac{a_1^2+c_1^2+a_2^2+c_2^2}{2} & 0 & 0 \\0& 0 & \frac{a_1^2+c_1^2+a_2^2+c_2^2}{2} & 0 \\0& 0 & 0 & a_2^2+c_2^2 \end{bmatrix},\quad i\in \mathbb{Z}.$$

Using the first equation on \eqref{newstieltjesORTH} with $A_{-1}=A$ and $C_0=C,$ we obtain
$$
B(z;W_{11})=
\begin{bmatrix}
\frac{\sqrt{(z-a_1^2-c_1^2)^2-4a_1^2c_1^2}}{(z-a_1^2-c_1^2)^2-4a_1^2c_1^2}& 0 & 0 &0 \\
0& * & 0 & 0\\
0& 0 & * & 0\\
0& 0 & 0 & \frac{\sqrt{(z-a_2^2-c_2^2)^2-4a_2^2c_2^2}}{(z-a_2^2-c_2^2)^2-4a_2^2c_2^2}
\end{bmatrix},
$$
where we used $dW_+=dW_-=d\tilde{\Sigma}(x),$ $\;d\tilde{\Sigma}(x)$ being the weight matrix given by \eqref{exTilde}.

It is easily seen that
$$
\lim_{z\uparrow 0}\frac{\sqrt{(z-a_k^2-c_k^2)^2-4a_k^2c_k^2}}{(z-a_k^2-c_k^2)^2-4a_k^2c_k^2}=\infty\;\;\Leftrightarrow\;a_k=c_k,\quad k=1,2,
$$
therefore, for $\rho=\begin{bmatrix}a & b \\b^* & 1-a\end{bmatrix},$ we obtain that
\begin{itemize}
  \item $a_1=c_1$ and $a_2=c_2$ implies that the walk is recurrent;
  \item $a_1\neq c_1$ and $a_2\neq c_2$ implies that the walk is transient;
  \item $a_1=c_1$ and $a_2\neq c_2$ implies that the walk is $\rho$-transient for $a=0$ and $\rho$-recurrent for $a>0$;
  \item $a_1\neq c_1$ and $a_2= c_2$ implies that the walk is $\rho$-transient for $a=1$ and $\rho$-recurrent for $a<1$.
\end{itemize}

{\color{black}
We observe that the walker returns infinitely often, in mean, to vertex $\ket{0}$ for any initial density operator when $a_1=c_1$ and $a_2=c_2$. In the contrasting scenario, where $a_j\neq c_j,\;j=1,2$, the walker returnees is finite in mean. Lastly, when only one of the values $a_j$ equals $c_j$, then the walk returns a finite number of times to $\ket{0}$, in mean, for one only density, and infinitely often for all others.

}

Moreover, the weight $dW_{11}$ is obtained by applications of the Perron-Stieltjes inversion formula:
$$
dW_{11}(x)=
\begin{bmatrix}
\left[\frac{\sqrt{(x-a_1^2-c_1^2)^2-4a_1^2c_1^2}}{(x-a_1^2-c_1^2)^2-4a_1^2c_1^2}\right]_+& 0 & 0 &0 \\
0& * & 0 & 0\\
0& 0 & * & 0\\
0& 0 & 0 & \left[\frac{\sqrt{(x-a_2^2-c_2^2)^2-4a_2^2c_2^2}}{(x-a_2^2-c_2^2)^2-4a_2^2c_2^2}\right]_+
\end{bmatrix}.
$$

\subsection{The case of simultaneous unitarily diagonalizable transitions} The above analysis can be applied to the simultaneous unitary diagonalizable coins, that is, we can take a unitary matrix $U$ and coins given by
$$
A=U\begin{bmatrix}
     a_1 & 0 \\
     0 & a_2
   \end{bmatrix}U^*,\;\;\;
C=U\begin{bmatrix}
     c_1 & 0 \\
     0 & c_2
   \end{bmatrix}U^*,\;\;\; a_1,a_2,c_1,c_2>0
$$
to obtain analogous conclusions about the recurrence of vertex $\ket{0}.$ In this case, we have
\begin{itemize}
  \item $a_1\leq c_1$ and $a_2\leq c_2\;\Rightarrow$ vertex $\ket{0}$ is recurrent;
  \item $a_1\leq c_1$ and $a_2> c_2\;\Rightarrow$ vertex $\ket{0}$ is $U\ket{e_1}\bra{e_1}U^*$-transient and $\rho$-recurrent for $\rho\neq U\ket{e_1}\bra{e_1}U^*;$
  \item $a_1> c_1$ and $a_2\leq c_2\;\Rightarrow$ vertex $\ket{0}$ is $U\ket{e_0}\bra{e_0}U^*$-transient and $\rho$-recurrent for $\rho\neq U\ket{e_0}\bra{e_0}U^*;$
  \item $a_1> c_1$ and $a_2>c_2\;\Rightarrow$ vertex $\ket{0}$ is transient.
\end{itemize}

\medskip

\begin{figure}[h!]
$$\begin{tikzpicture}[>=stealth',shorten >=1pt,auto,node distance=2cm]
    \node[state]         (q2) {$0$};
  \node[state]         (q3) [right of=q2]  {$1$};
\node[state]         (q4) [right of=q3]  {$2$};
\node[state]         (q5) [right of=q4]  {$3$};
\node         (q6) [right of=q5]  {$\ldots$};
  \path[->]          (q2)  edge         [bend left=15]     node {$A$}    (q3);
\path[->]          (q3)  edge         [bend left=15]     node {$A$}   (q4);
\path[->]          (q4)  edge         [bend left=15]       node {$A$} (q5);
  \path[->]          (q5)  edge         [bend left=15]     node {$C$}   (q4);
\path[->]          (q4)  edge         [bend left=15]      node {$C$}  (q3);
\path[->]          (q3)  edge         [bend left=15]      node {$C$}  (q2);
\path[->]          (q5)  edge         [bend left=15]      node {$A$}  (q6);
\path[->]          (q6)  edge         [bend left=15]    node {$C$}    (q5);
\draw [->] (q2) to[loop above] node[auto] {$B_{0}$} (q2);
\end{tikzpicture}$$
\caption{A slight modification on the first vertex.}
\label{graphB0}
\end{figure}
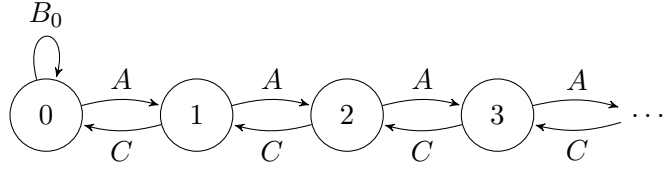

Let us describe an example of this and, in addition, let us consider a perturbation on the first vertex. In this case, the walk can be represented by Figure \ref{graphB0}, where $B_0$ represents the rate of jumping from vertex $\ket{0}$ to itself.

Let $U\in\mathbb{M}_2(\mathbb{C})$ be an unitary matrix and consider the CTOQW with generator
$$\hat{\mathcal{L}}=
\begin{bmatrix}
  G_0^\alpha+[B_0] & [C] &  &  &  \\
  [A] & G & [C] &  &  \\
  &[A] & G & [C] &   \\
   &  & \ddots & \ddots & \ddots
\end{bmatrix},\;A=C=U\begin{bmatrix}2 & 0\\0 & 1\end{bmatrix}U^*,\;B_0=U\begin{bmatrix}1 & h_1.i\\\overline{h_1}.i & 1\end{bmatrix}U^*,
$$
and the Hamiltonian operator
$$H=\sum_{i\in \mathbb{Z}_{\geq 0}}H_i\otimes\ket{i}\bra{i},\quad H_0=\begin{bmatrix}h_2 & h_1 \\\overline{h_1} & h_2\end{bmatrix}, h_2\in \mathbb{R},\;h_1\in\mathbb{C},\;H_i=0\mbox{ for }i>0.$$
The diagonal block matrices of $\hat{\mathcal{L}}$ are $G=-\mathcal{U}\textmd{diag}(8,5,5,2)\mathcal{U}^*$ and
$$G_0^\alpha=\mathcal{U}\begin{bmatrix}
        -4-|h_1|^2   & 0 & 0 & |h_1|^2\\
        0   & -5/2-|h_1|^2 & h_1^2 &0 \\
          0 & \overline{h_1}^2 & -5/2-|h_1|^2 &0 \\
          |h_1|^2 & 0 & 0 & -1-|h_1|^2
        \end{bmatrix}\mathcal{U}^*,$$
where $\mathcal{U}=U\otimes U,$ thus $G^\alpha_{0}$ is hermitian.

The Stieltjes Transform of the matrix weight associated to $\tilde{\mathcal{L}}$ ($\hat{\mathcal{L}}$ with $G_0^\alpha+[B_0]$ switched by $G$) is then, by Equation \eqref{Bs1},
\begin{equation}\label{BTilde-}
B(z,\tilde{\Sigma})=\frac{1}{32}\mathcal{U}
\begin{bmatrix}w_1(z)&0&0&0\\0&w_2(z)&0&0\\0&0&w_3(z)&0\\0&0&0&w_4(z)\end{bmatrix}\mathcal{U}^*,
\begin{array}{rl}
w_1(z)=   & 8-z-\sqrt{z(z-16)} \\
w_2(z)=& w_3(z)=20-4z-4\sqrt{z^2-10z+9}\\
w_4(z)=&32-16z-16\sqrt{z^2-4z}
\end{array}.
\end{equation}

The Stietjes Transform of $\hat{\mathcal{L}}$ is obtained by
\begin{equation}\nonumber
B(z,\Sigma)= \left(B(z,\tilde{\Sigma})^{-1}+(G_0^\alpha+[B_0]-G)\right)^{-1}
= \mathcal{U}
\begin{bmatrix}
  s_1(z) & 0 & 0 & |h_1|^2\\
  0 & s_2(z) & h_1^2 & 0\\
  0 & \bar{h_1}^2 & s_2(z) & 0\\
  |h_1|^2 & 0 & 0 & s_3(z)\\
\end{bmatrix}^{-1}\mathcal{U}^*,
\end{equation}
\begin{equation}\nonumber
\begin{split}
s_1(z)=& \frac{32}{z-8+\sqrt{z(z-16)}}+4-|h_1|^2,\;s_2(z)=\frac{8}{z-5+\sqrt{z^2-10z_9}}+\frac{5}{2}-|h_1|^2, \\
s_3(z)=&\frac{2}{z-2+\sqrt{z(z-4)}}+1-|h_1|^2.
\end{split}
\end{equation}

After some calculus using the limit given on Corollary \ref{CorRecurrence}, we obtain that this walk is recurrent for any choices of $h_1\in\mathbb{C},h_2\in \mathbb{R}.$

\medskip

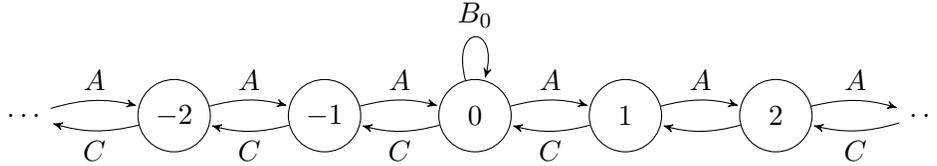
\begin{figure}[h!]
$$\begin{tikzpicture}[>=stealth',shorten >=1pt,auto,node distance=2cm]
\node[state] (q-2)      {$-2$};
\node[state]         (q-1) [right of=q-2]  {$-1$};
\node[state]         (q0) [right of=q-1]  {$0$};
\node[state]         (q1) [right of=q0]  {$1$};
\node[state]         (q2) [right of=q1]  {$2$};
\node         (qr) [right of=q2]  {$\ldots$};
\node         (ql) [left of=q-2]  {$\ldots$};
\path[->]          (ql)  edge         [bend left=15]  node[auto] {$A$}     (q-2);
\path[->]          (q-2)  edge         [bend left=15]   node[auto] {$A$}     (q-1);
\path[->]          (q-1)  edge         [bend left=15]   node[auto] {$A$}     (q0);
\path[->]          (q0)  edge         [bend left=15]   node[auto] {$A$}     (q1);
\path[->]          (q1)  edge         [bend left=15]   node[auto] {$A$}     (q2);
\path[->]          (q2)  edge         [bend left=15]   node[auto] {$A$}     (qr);
\path[->]          (qr)  edge         [bend left=15] node[auto] {$C$}       (q2);
\path[->]          (q2)  edge         [bend left=15]   node[auto] {$$}     (q1);
\path[->]          (q1)  edge         [bend left=15]  node[auto] {$C$}      (q0);
\path[->]          (q0)  edge         [bend left=15]   node[auto] {$C$}     (q-1);
\path[->]          (q-1)  edge         [bend left=15]  node[auto] {$C$}      (q-2);
\path[->]          (q-2)  edge         [bend left=15]  node[auto] {$C$}      (ql);
\draw [->] (q0) to[loop above]node[auto] {$B_{0}$}  (q0);
\end{tikzpicture}$$
\caption{Generator $\hat{\mathcal{L}}$ of a CTOQW on $\mathbb{Z}$ with a perturbation on vertex $\ket{0}$.}
\label{ZPerturbation}
\end{figure}

\textbf{A perturbation on the vertex $\ket{0}$ for the CTOQW on $\mathbb{Z}:$} We consider a CTOQW on $\mathbb{Z}$ with the same transitions as above but with a perturbation on vertex $\ket{0}.$ That is, we are taking the walk given by Figure \ref{ZPerturbation}, where
$$A=C=U\begin{bmatrix}2 & 0\\0 & 1\end{bmatrix}U^*,\quad B_0=U\begin{bmatrix}b_1 & b_2\\b_2 & b_3\end{bmatrix}U^*,\quad b_1,b_2,b_3\in\mathbb{R}.$$

{\color{black}
Each position on the lattice behaves similarly, except $\ket{0}$. The perturbation, characterized by a different matrix rate for a self-loop at $\ket{0}$, introduces a localized influence on the walker's behavior. This can be seen as a quantum interference effect, where the perturbation disrupts the otherwise uniform evolution of the quantum state.

Physically, this setup could be implemented in a quantum system where the different vertices of the integer lattice correspond to distinct quantum states, and the perturbation arises from a modification in the local dynamics at one specific position. This might be achieved through controlled interactions or external fields acting on the quantum system. Such perturbations can be leveraged in quantum algorithms and simulations, providing a way to encode specific information or perform quantum operations selectively at certain positions in the lattice.
}

We want to apply Equation \eqref{newstieltjes} to verify if vertex $\ket{0}$ is recurrent. To do this, we notice that
$$
\breve{\mathcal{L}}^+=
\begin{bmatrix}
G^\alpha_{0}+[B_{0}] & [C] &  & & \\
  [A]&G^\alpha & [C]   &&  \\
  &[A]&G^\alpha & [C]    & \\
   &  & \ddots &\ddots &\ddots
\end{bmatrix},\quad
\breve{\mathcal{L}}^-=
\begin{bmatrix}
  \ddots & \ddots & \ddots & & \\
  &[A] & G^\alpha  &[C]&  \\
  &&[A]& G^\alpha&[C] \\
   &  &  &[A] &G^\alpha
\end{bmatrix},
$$
where $G^\alpha=-\mathcal{U}\textmd{diag}(8,5,5,2)\mathcal{U}^*$ and now
\begin{equation}\nonumber
\begin{split}
G^\alpha_{0}=& -\frac{1}{2}\left[\left(A^*A+B_0^*B_0+C^*C\right)\otimes I+I\otimes\left(\overline{A^*A+B_0^*B_0+C^*C}\right)\right]-iH_0\otimes I+iI\otimes \overline{H_0} \\
     =& \mathcal{U}\begin{bmatrix}
        -8-|h_1|^2   & 0 & 0 & |h_1|^2\\
        0   & -5-|h_1|^2 & h_1^2 &0 \\
          0 & \overline{h_1}^2 & -5-|h_1|^2 &0 \\
          |h_1|^2 & 0 & 0 & -2-|h_1|^2
        \end{bmatrix}\mathcal{U}^*,
\end{split}
\end{equation}
which is a hermitian matrix.

The Stieltjes Transform of the matrix weight associated to $\breve{\mathcal{L}}^-$ is given on Equation \eqref{BTilde-}(since $A=C$) while the Stietjes Transform of the matrix weight associated to $\breve{\mathcal{L}}^+$ is
\begin{equation}\nonumber
\begin{split}
B(z,W_+)=& \left(B(z,W_-)^{-1}+(G^\alpha_0+[B_0]-G^\alpha)\right)^{-1} \\
=& \mathcal{U}\begin{bmatrix}
  \psi_1(z) & 0 & 0 & |h_1|^2\\
  0 & \psi_2(z) & h_1^2 & 0\\
  0 & \bar{h_1}^2 & \psi_2(z) & 0\\
  |h_1|^2 & 0 & 0 & \psi_3(z)\\
\end{bmatrix}^{-1}\mathcal{U}^*,
\end{split}
\end{equation}
where
\begin{equation}\nonumber
\begin{split}
\psi_1(z)=& \frac{32}{z-8+\sqrt{z(z-16)}}-|h_1|^2,\;\psi_2(z)=\frac{8}{z-5+\sqrt{z^2-10z_9}}-|h_1|^2, \\
\psi_3(z)=&\frac{2}{z-2+\sqrt{z(z-4)}}-|h_1|^2.
\end{split}
\end{equation}

Some calculus shows that
$$
-\lim_{z\uparrow 0}\textmd{Tr}\left(B(z,W_{11})\rho\right)=\infty
$$
for any choice of $h_1\in\mathbb{C},\;h_2\in \mathbb{R}$ and $\rho\in\mathbb{M}_2(\mathbb{C}),$ therefore vertex $\ket{0}$ is always recurrent for this CTOQW.

The same can be done with vertex $\ket{-1},$ however on this case we have to evaluate $-\lim_{z\uparrow 0}\textmd{Tr}\left(B(z,W_{22})\rho\right)=\infty,$ which is always infinite for any choice of $h_1\in\mathbb{C},\;h_2\in \mathbb{R}$ and $\rho\in\mathbb{M}_2(\mathbb{C}),$ therefore vertex $\ket{-1}$ is also always recurrent for this CTOQW.

\subsection{Noncommuting transitions}\label{NonC}

Let
$$
A=\begin{bmatrix}
    1 & 0 \\
    1 & -1
  \end{bmatrix},\;\;
C=\begin{bmatrix}
    1 & 1 \\
    0 & -1
  \end{bmatrix},
$$
where
$$G_1=-3I_4,\;\;
G_0=\frac{1}{2}\begin{bmatrix}
      -4 & 1 & 1 & 0 \\
      1 & -3 & 0 & 1 \\
      1 & 0 & -3 & 1 \\
      0 & 1 & 1 & -2
    \end{bmatrix},\;
G_2=-\frac{1}{2}\begin{bmatrix}
      2 & 1 & 1 & 0 \\
      1 & 3 & 0 & 1 \\
      1 & 0 & 3 & 1 \\
      0 & 1 & 1 & 4
    \end{bmatrix}.
$$

Consider the CTOQW with $V=\{0,1,2,3\}$ induced by the generator
$$
\hat{\mathcal{L}}=
\begin{bmatrix}
  G_0 & \lceil C\rceil &0& 0\\
  \lceil A\rceil  & G_1 &\lceil C\rceil &0 \\
  0&\lceil A\rceil  & G_1 &\lceil C\rceil  \\
  0&0 & \lceil A\rceil  & G_2
\end{bmatrix}.
$$

Note that this generator satisfies the conditions \eqref{EqDette} with $R_n=I_4,\;n=0,1,2,3,$ thus there exists a positive weight matrix associated to $\hat{\mathcal{L}},$ which will be evaluated now.

The eigenvalues of $-\hat{\mathcal{L}}$ are
\begin{equation}\nonumber
\begin{split}
\lambda_1=& 0,\;\lambda_2=3-\sqrt{5},\;\lambda_3=3+\sqrt{5},\;\lambda_4=3-\sqrt{7},\;\lambda_5=3+\sqrt{7}, \\
\lambda_6=&\frac{7-\sqrt{17}}{4},\;\lambda_7=\frac{7+\sqrt{17}}{4},\;\lambda_8=\frac{11-\sqrt{41}}{4},\;\lambda_9=\frac{11+\sqrt{41}}{4},
\end{split}
\end{equation}
($\lambda_1,\lambda_4,\lambda_5,\lambda_6,\lambda_7,\lambda_8$ and $\lambda_9$ have multiplicity 2) with weights
\begin{equation}\nonumber
\begin{split}
 W_1=   & \frac{1}{20}\begin{bmatrix}
      3 & -1 & -1 & 2 \\
      -1 & 2 & 2& 1 \\
      -1 & 2 & 2 & 1 \\
      2 & 1 & 1 & 3
    \end{bmatrix},\;
W_2=\frac{1}{2}\left(W_1+\frac{\sqrt{5}}{20}Y\right),\;W_3=\frac{1}{2}\left(W_1-\frac{\sqrt{5}}{20}Y\right),\\
W_4=&\frac{1}{14}\left((14+3\sqrt{7})W_1+\frac{\sqrt{7}}{4}Y\right),\;W_5=\frac{1}{14}\left((14-3\sqrt{7})W_1-\frac{\sqrt{7}}{4}Y\right),\\
W_6=&\frac{1}{4}\left(1+\frac{\sqrt{17}}{17}\right)\left(I_4-4W_1\right),W_7=\frac{1}{4}\left(1-\frac{\sqrt{17}}{17}\right)\left(I_4-4W_1\right)\\
W_8=&\frac{1}{4}\left(1+\frac{4\sqrt{41}}{41}\right)\left(I_4-4W_1\right),W_9=\frac{1}{4}\left(1-\frac{4\sqrt{41}}{41}\right)\left(I_4-4W_1\right).
\end{split}
\end{equation}
where
$$
Y=\begin{bmatrix}-1 & 1 & 1 & 0\\1 & 0& 0 & 1\\1 & 0 & 0 & 1\\0 & 1 & 1 & 1\end{bmatrix}.
$$

For instance, we have for $\rho=\begin{bmatrix}a & b \\b^* & 1-a\end{bmatrix},$
$$p_{00;\rho}(t) = \sum_{k=1}^{9}e^{-\lambda_k}W_k=\frac{1}{4}+\left(e^{-\lambda_2t}-e^{-\lambda_3t}\right)v_1+\frac{e^{-\lambda_2t}+e^{-\lambda_3t}}{8}+
\left(e^{-\lambda_4t}-e^{-\lambda_5t}\right)v_2+\frac{e^{-\lambda_4t}+e^{-\lambda_5t}}{4},
$$
where $v_1=\dfrac{\sqrt{5}}{40}(1-2a+4\textmd{Re}(b))$ and $v_2=\dfrac{\sqrt{7}}{28}(2-a+2\textmd{Re}(b)).$

\begin{figure}[!ht]
\centering
\includegraphics[width=0.6\textwidth]{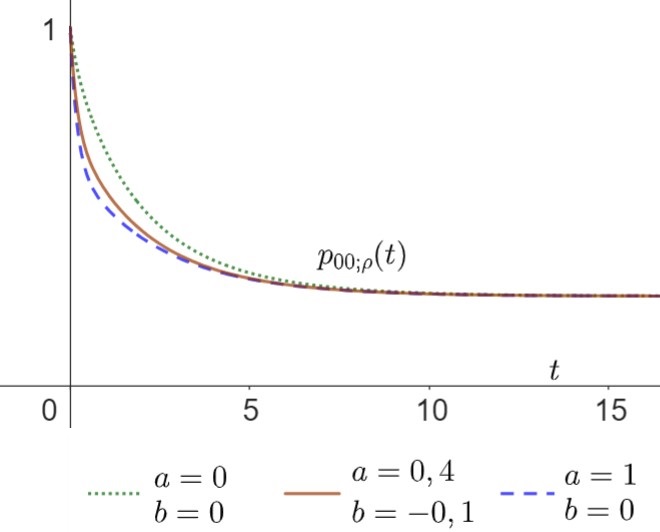}
\caption{Different transition probabilities for Example \ref{NonC}.}
\label{decoherence}
\end{figure}

{\color{black} Non-diagonal matrices often lead to more intricate transition probabilities in CTOQWs. Therefore, these distinct values introduce more complex interference effects and allow the walker's behavior dependence on $b$. Furthermore, the variation of the initial density distribution can be understood as a conceptual analogy for decoherence. Initially, the uniform distribution between quantum states exemplifies a coherent starting point. However, the introduction of decoherence involves changing this initial distribution, with the subsequent evolution of CTOQW reflecting the influence of this perturbation, similar to the loss of coherence observed in decoherence phenomena. This change in initial density alters the probabilistic evolution, illustrating the sensitivity of the quantum system to its initial conditions. See Figure \ref{decoherence} to compare the transition probabilities for different initial densities.}

\subsection{Antidiagonal transitions: another approach}

Let us discuss an example with antidiagonal transitions. We do this in terms of preliminary reasoning with a generator that has alternating matrices. More precisely, we consider a block matrix of the form
\begin{equation*}\label{periodL}
J=\begin{bmatrix}
-G_0&\lceil P_0\rceil &&&  &  & \\ \lceil P_0\rceil &-G&\lceil P_1\rceil &  &  &  &\\ &\lceil P_1\rceil &-G&\lceil P_0\rceil &  &  &  &\\ &&\lceil P_0\rceil &-G&\lceil P_1\rceil &&\\ &&&\lceil P_1\rceil &-G&\lceil P_0\rceil &  & \\&&&&&\ddots &\ddots&\ddots
  \end{bmatrix},
\end{equation*}
where
$$P_0=\begin{bmatrix}\sqrt{a_2c_1} & 0\\0&\sqrt{a_1c_2} \end{bmatrix},\;\;P_1=\begin{bmatrix}\sqrt{a_1c_2} & 0\\0&\sqrt{a_2c_1} \end{bmatrix},\;\;a_1,a_2,c_1,c_2\geq 0,$$
$$
G_0=-\begin{bmatrix}a_2^2 &0&0&0\\0&\frac{a_1^2+a_2^2}{2}&0&0\\0&0&\frac{a_1^2+a_2^2}{2}&0\\0&0&0& a_1^2\end{bmatrix},\;
G=-\begin{bmatrix}a_2^2+c_2^2&0&0&0\\0&\frac{a_1^2+a_2^2+c_1^2+c_2^2}{2}&0&0\\0&0&\frac{a_1^2+a_2^2+c_1^2+c_2^2}{2}&0\\0&0&0& a_1^2+c_1^2\end{bmatrix}.
$$

We notice that $J$ may not be a valid generator of a CTOQW, however this block matrix will be auxiliary to obtain a weight matrix associated to a specific kind of generator later. Then, we use \cite[Theorem 2.4]{detteC} to obtain the following equality associated to the Stieltjes transform of the weight matrix $d\Sigma(x)$ associated to $\tilde{J},$ which is the equivalent of $J$ with $G_0$ switched by $G:$
$$
B(z,\tilde{\Sigma})=\{zI_4-G+\lceil P_0\rceil \{zI_4-G+\lceil P_1\rceil B(z,\tilde{\Sigma})(-\lceil P_1\rceil )\}^{-1}(-\lceil P_0\rceil )\}^{-1},
$$
where $R_i=I_4$ for every $R_i$ appearing on \cite[Theorem 2.4]{detteC} is a consequence of $\lceil P_0\rceil =\lceil P_0\rceil ^T$ and $\lceil P_1\rceil =\lceil P_1\rceil ^T.$

The known matrices of the equality are all diagonal, thus we assume that $$B(z,\tilde{\Sigma})=\textmd{diag}\left(\tilde{f}_1(z),\tilde{f}_2(z),\tilde{f}_3(z),\tilde{f}_4(z)\right),$$
and then each $\tilde{f}_k(z)$ is a solution of
$$
\tilde{f}_k(z)=\{z-\tilde{g}_k-m_{0,k}\{z-\tilde{g}_k-m_{1,k}f_k(z)m_{1,k}\}^{-1}m_{0,k}\}^{-1},
$$
where $G=\textmd{diag}(\tilde{g}_1,\tilde{g}_2,\tilde{g}_3,\tilde{g}_4),\;\;\lceil P_j\rceil =\textmd{diag}(m_{j,1},m_{j,2},m_{j,3},m_{j,4}),\;j=0,1.$
Some algebra gives
$$
m_{1,k}^2(z-\tilde{g}_k)\tilde{f}_k(z)^2+(m_{0,k}^2-m_{1,k}^2-(z-\tilde{g}_k)^2)\tilde{f}_k(z)+(z-\tilde{g}_k)=0.
$$
Therefore
$$
\tilde{f}_k(z)=\frac{m_{1,k}^2-m_{0,k}^2+(z-\tilde{g}_k)^2- \sqrt{\left(m_{0,k}^2-m_{1,k}^2-(z-\tilde{g}_k)^2\right)^2-4(z-\tilde{g}_k)^2m_{1,k}^2}}{2m_{1,k}^2(z-\tilde{g}_k)}.
$$

As usual, the next step is to obtain the Stieltjes transform of $d\Sigma,$ the weight matrix associated to $J.$ By equation (2.20) of \cite{detteC}, we have
$$
B(z,\Sigma)= \left(B(z,\tilde{\Sigma})^{-1}+(G_0-G)\right)^{-1}=\mathrm{diag}(f_1(z),f_2(z),f_3(z),f_4(z)),
$$
where
$$
f_k(z)=\frac{1}{2}\frac{\psi_k(z)m_{1,k}-m_{1,k}\sqrt{\psi_k(z)^2+4\gamma_k(z)^2}-2\tilde{g}_k\gamma_k(z)+2g_{k}\gamma_k(z)}{m_{1,k}^2\gamma_k(z)-\tilde{g}_k^2\gamma_k(z)-g_{k}^2\gamma_k(z)+2\tilde{g}_kg_{k}\gamma_k(z)-m_{1,k}g_{k}\psi_k(z)+m_{1,k}\tilde{g}_k\psi_k(z)},
$$
and we have put $G_0=\mathrm{diag}(g_1,g_2,g_3,g_4,)$ $\psi_k(z)=-(z+g_k)^2+m_{1,k}^2-m_{0,k}^2,\;\;\gamma_k(z)=(z+g_k)m_{1,k}.$

\medskip

Now, we are able to consider an antidiagonal transition in the following terms: consider a CTOQW on $\mathbb{Z}_+$ whose generator is of the form
$$\hat{\mathcal{L}}=
\begin{bmatrix}
  G_0 & \lceil C\rceil  &  &  &  \\
  \lceil A\rceil  & G & \lceil C\rceil  &  &  \\
  &\lceil A\rceil  & G & \lceil C\rceil  &   \\
   &  & \ddots & \ddots & \ddots
\end{bmatrix},\;
A=\begin{bmatrix}0 & a_1 \\a_2 & 0\end{bmatrix},\;C=\begin{bmatrix}0 & c_1 \\c_2 & 0\end{bmatrix},$$
$$
G_0=-\begin{bmatrix}a_2^2 &0&0&0\\0&\frac{a_1^2+a_2^2}{2}&0&0\\0&0&\frac{a_1^2+a_2^2}{2}&0\\0&0&0& a_1^2\end{bmatrix},\;
G=-\begin{bmatrix}a_2^2+c_2^2&0&0&0\\0&\frac{a_1^2+a_2^2+c_1^2+c_2^2}{2}&0&0\\0&0&\frac{a_1^2+a_2^2+c_1^2+c_2^2}{2}&0\\0&0&0& a_1^2+c_1^2\end{bmatrix}.
$$

We have the symmetrization
$$
J=R(-\hat{\mathcal{L}})R^{-1}=\begin{bmatrix}
-G_0&\lceil P_0\rceil &&&  &  & \\ \lceil P_0\rceil &-G&\lceil P_1\rceil &  &  &  &\\ &\lceil P_1\rceil &-G&\lceil P_0\rceil &  &  &  &\\ &&\lceil P_0\rceil &-G&\lceil P_1\rceil &&\\ &&&\lceil P_1\rceil &-G&\lceil P_0\rceil &  & \\&&&&&\ddots &\ddots&\ddots
  \end{bmatrix},\;
R=\mathrm{diag}(\lceil R_0\rceil ,\lceil R_1\rceil ,\ldots),
$$
where
$$
R_{2k}=\begin{bmatrix}
         \left(\dfrac{c_2}{a_1}\right)^{\frac{k}{2}}\left(\dfrac{c_1}{a_2}\right)^{\frac{k-2}{2}} & 0 \\
         0 & \left(\dfrac{c_2}{a_1}\right)^{\frac{k-2}{2}}\left(\dfrac{c_1}{a_2}\right)^{\frac{k}{2}}
       \end{bmatrix},\;\;
R_{2k+1}=\left(\frac{c_1c_2}{a_1a_2}\right)^{\frac{k}{2}}\begin{bmatrix}0 & 1 \\1 & 0\end{bmatrix},\;\;k=0,1,2,\ldots,
$$
and $P_0$ and $P_1$ are the ones given above. Thus $J$ and $\hat{\mathcal{L}}$ have the same associated weight matrix and we obtain, for $d\Sigma(x)$ given above that
$$
\lim_{z\uparrow 0}\textmd{Tr}\left(B(z,\Sigma)\rho\right)=\lim_{z\uparrow 0}(f_1(z)a+f_4(z)(1-a)),
$$
where $\rho=\begin{bmatrix}a & b \\b^* & 1-a\end{bmatrix}.$ After some calculus we obtain that
\begin{equation}\nonumber
\begin{split}
\lim_{z\uparrow 0}f_1(z)=\infty\;\Leftrightarrow&\;\;a_1=\sqrt{\frac{2c_2^4-a_2^2c_1^2+a_2^4+3a_2^2c_2^2}{a_2^2+2c_2^2}},\;2c_2^4+a_2^4+3a_2^2c_2^2>a_2^2c_1^2,  \\
\lim_{z\uparrow 0}f_4(z)=\infty\;\Leftrightarrow&\;\;a_2=\sqrt{\frac{2c_1^4-a_1^2c_2^2+a_1^4+3a_1^2c_1^2}{a_1^2+2c_1^2}},\;2c_1^4+a_1^4+3a_1^2c_1^2>a_1^2c_2^2,
\end{split}
\end{equation}
giving the following conclusion (see Corollary \ref{CorRecurrence}):
\begin{itemize}
\item $a_1=\sqrt{\frac{2c_2^4-a_2^2c_1^2+a_2^4+3a_2^2c_2^2}{a_2^2+2c_2^2}}$ and $a_2=\sqrt{\frac{2c_1^4-a_1^2c_2^2+a_1^4+3a_1^2c_1^2}{a_1^2+2c_1^2}}\Rightarrow$ vertex $\ket{0}$ is recurrent;
  \item $a_1=\sqrt{\frac{2c_2^4-a_2^2c_1^2+a_2^4+3a_2^2c_2^2}{a_2^2+2c_2^2}}$ and $a_2\neq\sqrt{\frac{2c_1^4-a_1^2c_2^2+a_1^4+3a_1^2c_1^2}{a_1^2+2c_1^2}}\Rightarrow$ vertex $\ket{0}$ is $\rho$-transient when $a=0$ and $\rho$-recurrent when $a>0;$
  \item $a_1\neq\sqrt{\frac{2c_2^4-a_2^2c_1^2+a_2^4+3a_2^2c_2^2}{a_2^2+2c_2^2}}$ and $a_2=\sqrt{\frac{2c_1^4-a_1^2c_2^2+a_1^4+3a_1^2c_1^2}{a_1^2+2c_1^2}}\Rightarrow$ vertex $\ket{0}$ is $\rho$-transient when $a=1$ and $\rho$-recurrent when $a<1;$
  \item $a_1\neq\sqrt{\frac{2c_2^4-a_2^2c_1^2+a_2^4+3a_2^2c_2^2}{a_2^2+2c_2^2}}$ and $a_2\neq\sqrt{\frac{2c_1^4-a_1^2c_2^2+a_1^4+3a_1^2c_1^2}{a_1^2+2c_1^2}}\Rightarrow$ vertex $\ket{0}$ is transient.
\end{itemize}

{\color{black}
Hence, as in Example \ref{exDiag}, the recurrence only depends on the density once the values of A and C are defined. Although the conclusion is similar to the case in Example \ref{exDiag}, the more complex values above arise from the representations of $\lceil P_0\rceil$ and $\lceil P_1\rceil$, which are not diagonal.

\section{Appendix}\label{apendice}
In this appendix, we will recall some well-known results from the theory of matrix orthogonal polynomials, underlining their relevance and application in the context of this work.

\medskip

1. Let $\Sigma$ be a $d^2\times d^2$ weight matrix and denote by
$$
S_k=\int x^kd\Sigma(x),\quad k=0,1,\ldots
$$
the corresponding moments. The block Hankel matrices are defined by
$$
\underline{H}_{2m}=
\begin{bmatrix}
  S_0 & \cdots & S_m \\
  \vdots &  & \vdots \\
  S_m & \cdots & S_{2m}
\end{bmatrix},\;\;\; m\geq 0.
$$

\begin{teo}[Slight adaptation of Theorem 2.1 \cite{detteC}]\label{teoDette}
Consider the block matrix $\hat{\mathcal{L}}$ given by Equation \eqref{CTQtridiZ+}, assume that $A_n,C_{n+1},\;n\geq 0,$ are nonsingular matrices and $B_n\leq 0\;\forall n$. Now let $\{Q_n(x)\}_{n\geq0}$ be the sequence of matrix-valued polynomials defined by \eqref{CT3thermZ+}. Then there exists a weight matrix $\Sigma$ with positive definite block Hankel matrices $\underline{H}_{2m}, m\geq 0$, such that the sequence of polynomials $\{Q_n(x)\}_{n\geq0}$ is orthogonal with respect to $\Sigma$ if and only if there is a sequence of nonsingular matrices $(R_n)_{n\geq0}$ such that
\begin{equation}\label{EqDette}
\begin{split}
&R_nB_nR_n^{-1}\mbox{ is hermitian,}\quad n\geq 0,   \\
&R_n^*R_n=(A_0^*\cdots A_{n-1}^*)^{-1}R_0^*R_0C_1\cdots C_n,\quad n\geq 0.
\end{split}
\end{equation}
Moreover, $S_0=\left(R_0^*R_0\right)^{-1}.$
\end{teo}

\medskip

2. Perturbation of the Stieltjes transform:

\begin{teo}[Theorem 2.3 of \cite{detteC}]
Consider the block matrix $\hat{\mathcal{L}}$ given by Equation \eqref{CTQtridiZ+} and the matrix $\tilde{\mathcal{L}}$ which is the same as $\hat{\mathcal{L}}$ but with a perturbation on the first block, that is,
\begin{equation*}\label{CTQtridiZTilde}
\tilde{\mathcal{L}}=\begin{bmatrix}
                \tilde{B}_0 & C_1 &  &  &  \\
                A_0 & B_1 & C_2 &  &  \\
                 & A_1 & B_2 &  C_3&  \\
                 &  & \ddots & \ddots & \ddots
              \end{bmatrix}.
\end{equation*}
If $\Sigma$ is the weight matrix associated to $\hat{\mathcal{L}}$ with positive definite block Hankel matrices such that $R_0\tilde{B}_0R_0^{-1}$ is symmetric and such that $(R_n)_{n\geq0}$ is a sequence of matrices which satisfies condition \eqref{EqDette}, then there exists a weight matrix $\Sigma$ corresponding to $\tilde{\mathcal{L}}.$ If the weight matrix $\Sigma$ and $\tilde{\Sigma}$ are determined by their moments, then the Stieltjes transforms of the weights satisfy
\begin{equation*}\label{tildeEq}
B(z,\Sigma)=\left\{B(z,\tilde{\Sigma})^{-1}-S_0^{-1}\left(\tilde{B}_0-B_0\right)\right\}^{-1}.
\end{equation*}
\end{teo}

\medskip

3. Explicit weight matrix for a class of walks on the half-line. The following is a restatement of a result due to A.J. Dur\'an \cite{duran-ratio}:
let $A$ be positive definite and define
$$H(z)=A^{-1/2}(B-zI)A^{-1}(B-zI)A^{-1/2}-4I.$$
Such matrix is diagonalizable except for at most finitely many complex numbers $z$, so that we can write $-H(z)=U(z)D(z)U^{-1}(z)$, where $D(z)$ is a diagonal matrix with diagonal entries $\{d_{ii}(z)\}$. For $x$ real, we have that $-H(z)$ is Hermitian, so it is unitarily diagonalizable, that is, we can have $U(x)$ such that $U(x)U^*(x)=I$. Also, $D(z)$ has real entries. With such matrices defined, we have:

\medskip

\begin{teo}\cite{duran-ratio} If A is positive definite and B Hermitian, the weight matrix for the matrix-valued polynomials defined by
$$tU_n(t)=U_{n+1}(t)A+U_n(t)B+U_{n-1}(t)A,\;\;\;n\geq 0,\;\;\;U_0(t)=I,\;\;\;U_{-1}(t)=0,$$
is the matrix of weights given by
$$dW(x)=\frac{1}{2\pi}A^{-1/2}U(x)(D^+(x))^{1/2}U^*(x)A^{-1/2}dx,$$
where $D^+(z)$ is a diagonal matrix with diagonal entries $d_{ii}^+(z)=\max\{d_{ii}(x),0\}$.
\end{teo}

\medskip

}
\bigskip

{\bf Acknowledgements.} The author wishes to thank his Ph.D. advisor Carlos F. Lardizabal. The author is also grateful to M. Dom\'inguez de la Iglesia for interesting discussions around the topic. This work is financially supported by CAPES (Coordena\c c\~ao de Aperfei\c coamento de Pessoal de N\'ivel Superior) and Instituto Federal de Mato Grosso do Sul.

\medskip

\textbf{Declaration of competing interest}. There are no competing interests.

\textbf{Data Availability}. The datasets generated during the current study are available from the corresponding author on reasonable request.

\nocite{*}
\bibliographystyle{abnt}

\end{document}